\documentclass[a4paper,fleqn,usenatbib,longauth]{mnras} 

\usepackage{newtxtext,newtxmath} 
\usepackage[T1]{fontenc}
\usepackage{ae,aecompl}

\usepackage{graphicx}	
\usepackage{subfigure}
\usepackage{pdflscape} 
\usepackage{booktabs} 
\usepackage{multirow,bigdelim} 
\usepackage{multicol} 
\usepackage{ragged2e} 
\usepackage{siunitx} 
\usepackage{tabulary} 
\usepackage{tabularx} 
\newcolumntype{X}{>{\raggedright\arraybackslash}X} 
\newcolumntype{Y}{>{\centering\arraybackslash}X} 
\newcolumntype{Z}{>{\raggedleft\arraybackslash}X} 

\usepackage{verbatim} 

\newcommand{\kms}{km~s$^{-1}$} 
\newcommand{\msun}{$M_{\odot}$}

\newcommand{\mstar}{$M_{\ast}$}

\newcommand{\ms}{SFR-M$_{\ast}$}
\newcommand{\sfruv}{SFR$_{\rm UV}$}
\newcommand{\sfrir}{SFR$_{\rm IR}$}

\newcommand{\ntot}{193} 

\newcommand{\nscuba}{193}
\newcommand{\nrxa}{63}
\newcommand{\paperII}{Paper II}
\newcommand{\paperIII}{Paper III}
\newcommand{\paperiv}{Paper IV}
\newcommand{\numberCO}{90}
\newcommand{\edit}{ }

\title[JINGLE: survey overview]{JINGLE, a JCMT legacy survey of dust and gas for galaxy evolution studies: I. Survey overview and first results}

\author[Saintonge et al.]{\Large \parbox{\textwidth}{Am\'elie Saintonge$^{1}$\thanks{E-mail: a.saintonge@ucl.ac.uk}, 
Christine D. Wilson$^{2}$, 
Ting Xiao$^{3, 43}$, 
Lihwai Lin$^{4}$, 
Ho Seong Hwang$^{5}$, 
Tomoka Tosaki$^{6}$,
Martin Bureau$^{24}$,
Phillip J. Cigan$^{26}$,
Christopher J. R. Clark$^{26}$,
David L. Clements$^{27}$,
Ilse De Looze$^{1,16}$,
Thavisha Dharmawardena$^{4,20}$,
Yang Gao$^{43}$,
Walter K. Gear$^{26}$,
Joshua Greenslade$^{27}$,
Isabella Lamperti$^{1}$,
Jong Chul Lee$^{19}$,
Cheng Li$^{41}$,
Micha{\l} J.~Micha{\l}owski$^{23,36}$,  
Angus Mok$^{2}$,
Hsi-An Pan$^{4}$,
Anne E. Sansom$^{28}$,
Mark Sargent$^{25}$,
Matthew W. L. Smith$^{26}$,
Thomas Williams$^{26}$, 
Chentao Yang$^{9,10}$,
Ming Zhu$^{14}$,
Gioacchino Accurso$^{1}$,
Pauline Barmby$^{40}$, 
Nathan Bourne$^{23}$,
Elias Brinks$^{22}$,
Toby Brown$^{2}$, 
Aeree Chung$^{18}$,
Eun Jung Chung$^{19}$,
Anna Cibinel$^{25}$,
Kristen Coppin$^{22}$,
Jonathan Davies$^{26}$,
Timothy A. Davis$^{26}$,
Steve Eales$^{26}$,
Lapo Fanciullo$^{4}$,
Taotao Fang$^{8}$,
Yu Gao$^{9,10}$,
David H. W. Glass$^{28}$,
Haley L. Gomez$^{26}$,
Thomas Greve$^{1}$,
Jinhua He$^{11,12,13}$,
Luis C. Ho$^{34,35}$, Feng Huang$^{8}$,
Hyunjin Jeong$^{19}$,
Xuejian Jiang$^{9,10}$,
Qian Jiao$^{14}$,
Francisca Kemper$^{4}$,
Ji Hoon Kim$^{17}$,
Minjin Kim$^{19}$,
Taehyun Kim$^{19}$,
Jongwan Ko$^{19}$,
Xu Kong$^{15}$, 
Kevin Lacaille$^{2}$,
Cedric G. Lacey$^{29}$,
Bumhyun Lee$^{18}$,
Joon Hyeop Lee$^{19}$,
Wing-Kit Lee$^{4,21}$,
Karen Masters$^{30,44}$,
Se-Heon Oh$^{19}$,
Padelis Papadopoulos$^{26}$,
Changbom Park$^{5}$,
Sung-Joon Park$^{19}$,
Harriet Parsons$^{33}$ Kate Rowlands$^{31}$,
Peter Scicluna$^{4}$,
Jillian M. Scudder$^{25,39}$,
Ramya Sethuram$^{43}$,
Stephen Serjeant$^{32}$,
Yali Shao$^{34}$, Yun-Kyeong Sheen$^{19}$,
Yong Shi$^{37,38}$,
Hyunjin Shim$^{42}$,
Connor M. A. Smith$^{26}$, 
Kristine Spekkens$^{7}$,
An-Li Tsai$^{20}$,
Sheona Urquhart$^{32}$, 
Aprajita Verma$^{24}$,
Giulio Violino$^{22}$,
Serena Viti$^{1}$,
David Wake$^{32}$,
Junfeng Wang$^{8}$, Jan Wouterloot$^{33}$, 
Yujin Yang$^{19}$, Kijeong Yim$^{19}$,
Fangting Yuan$^{43}$ and Zheng Zheng$^{14}$} \\ 
\\
{\parbox{\textwidth}{\scriptsize$^{1}$Department of Physics and Astronomy, University College London, Gower Street, London, WC1E 6BT, UK\\
$^{2}$Department of Physics and Astronomy, McMaster University, Hamilton ON Canada L8S 4M1 \\
$^{3}$Department of Physics, Zhejiang University, Hangzhou, Zhejiang 310027, China \\
$^{4}$Institute of Astronomy and Astrophysics, Academia Sinica, P.O. Box 23-141, Taipei 10617, Taiwan \\
$^{5}$School of Physics, Korea Institute for Advanced Study, 85 Hoegiro, Dongdaemun-gu, Seoul 02455, Republic of Korea \\
$^{6}$Joetsu University of Education, Yamayashiki-machi, Joetsu, Niigata 943-8512, Japan \\
$^{7}$Department of Physics, Royal Military College of Canada, PO Box 17000, Station Forces, Kingston, ON K7K 7B4 \\
$^{8}$Department of Astronomy and Institute of Theoretical Physics and Astrophysics, Xiamen University, Xiamen, China 361005 \\
$^{9}$Purple Mountain Observatory, Chinese Academy of Sciences, Nanjing 210008, China\\
$^{10}$Key Laboratory of Radio Astronomy, Chinese Academy of Sciences, Nanjing 210008, China\\
$^{11}$Key Laboratory for the Structure and Evolution of Celestial Objects, Yunnan observatories, Chinese Academy of Sciences, P.O. Box 110, Kunming, 650011, Yunnan Province, China\\
$^{12}$Chinese Academy of Sciences, South America Center for Astrophysics (CASSACA), Camino El Observatorio 1515, Las Condes, Santiago, Chile\\
$^{13}$Departamento de Astronomia, Universidad de Chile, Casilla 36-D, Santiago, Chile\\
$^{14}$National Astronomical Observatory of China, 20A Datun Road, Chaoyang District, Beijing, China 100012 \\
$^{15}$Key Laboratory for Research in Galaxies and Cosmology, Department of Astronomy, University of Science and Technology of China, Hefei 230026, China\\
$^{16}$Sterrenkundig Observatorium, Universiteit Gent, Krijgslaan 281 S9, B9000 Gent, Belgium\\
$^{17}$Subaru Telescope, National Astronomical Observatory of Japan, 650 North A'ohoku Place, Hilo, HI, 96720, USA \\
$^{18}$Department of Astronomy, Yonsei University, 50 Yonsei-ro, Seodaemun-gu, Seoul, 03722, South Korea\\
$^{19}$Korea Astronomy and Space Science Institute, 776 Daedeokdae-ro, Yuseong-gu, Daejeon, 34055, South Korea\\
$^{20}$National Central University, No. 300, Zhongda Rd., Zhongli District, Taoyuan City 32001, Taiwan\\
$^{21}$Center for Interdisciplinary Exploration and Research in Astrophysics (CIERA) and Department of Physics and Astronomy, Northwestern University, 2145 Sheridan Road, Evanston, IL 60201, USA\\
$^{22}$Centre for Astrophysics Research, University of Hertfordshire, College Lane, AL10 9AB, UK \\
$^{23}$Scottish Universities Physics Alliance (SUPA), Institute for Astronomy, University of Edinburgh, Royal Observatory, Blackford Hill, Edinburgh, EH9 3HJ, UK \\
$^{24}$Sub-department of Astrophysics, University of Oxford, Denys Wilkinson Building, Keble Road, Oxford, OX1 3RH, UK\\
$^{25}$Astronomy Centre, Department of Physics and Astronomy, University of Sussex, Brighton BN1 9QH, England \\
$^{26}$School of Physics and Astronomy, Cardiff University, Queens Buildings, The Parade, Cardiff, CF24 3AA, UK \\
$^{27}$Blackett Laboratory, Physics Department, Imperial College, London, SW7 2AZ, UK \\
$^{28}$Jeremiah Horrocks Institute, University of Central Lancashire, Preston, Lancashire, PR1 2HE, UK \\
$^{29}$Institute for Computational Cosmology, Physics Dept., Durham University, South Road, Durham, DH1 3LE, UK \\
$^{30}$Institute of Cosmology and Gravitation, University of Portsmouth, Dennis Sciama Building, Portsmouth, PO1 3FX, UK \\
$^{31}$Department of Physics \& Astronomy, Johns Hopkins University, Bloomberg Center, 3400 N. Charles St., Baltimore, MD 21218, USA \\
$^{32}$School of Physical Sciences, The Open University, Walton Hall, Milton Keynes, MK7 6AA, UK \\
$^{33}$East Asian Observatory, 660 North A'Ohoku Place, Hilo, Hawaii, 96720, USA \\
$^{34}$Kavli Institute for Astronomy and Astrophysics, Peking University, Beijing 100871, China \\
$^{35}$Department of Astronomy, School of Physics, Peking University, Beijing 100871, China \\
$^{36}$Astronomical Observatory Institute, Faculty of Physics, Adam Mickiewicz University, ul.~S{\l}oneczna 36, 60-286 Pozna{\'n}, Poland \\
$^{37}$ Key Laboratory of Modern Astronomy and Astrophysics (Nanjing University), Ministry of Education, Nanjing 210093, China \\
$^{38}$School of Astronomy and Space Science, Nanjing University, Nanjing 210093, China \\
$^{39}$Department of Physics, Oberlin College, Oberlin, Ohio, 44074, USA \\
$^{40}$Department of Physics and Astronomy and Centre for Planetary Science and Exploration, University of Western Ontario, London,
ON, N6A 3K7, Canada\\
$^{41}$Tsinghua Center for Astrophysics and Physics Department, Tsinghua University, Beijing 100084, China \\
$^{42}$Department of Earth Science Education, Kyungpook National University, 80 Daehak-ro, Buk-gu, Daegu, 41556, South Korea \\
$^{43}$Shanghai Astronomical Observatory, 80 Nandan Road, Xuhui District, Shanghai, China 200030 \\
$^{44}$Department of Physics and Astronomy, Haverford College, 370 Lancaster Avenue, Haverford, PA 19041, USA }}}

\date{Accepted 2018 September 08. Received 2018 September 08; in original form 2017 September 25}

\pubyear{2018}

\begin{document}
\label{firstpage}
\pagerange{\pageref{firstpage}--\pageref{lastpage}}
\maketitle

\begin{abstract}
JINGLE is a new JCMT legacy survey designed to systematically study the cold interstellar medium of galaxies in the local Universe. As part of the survey we perform 850$\mu$m continuum measurements with SCUBA-2 for a representative sample of \ntot\ {\it Herschel}-selected galaxies with $M_{\ast}>10^9M_{\odot}$, as well as integrated CO(2-1) line fluxes with RxA3m for a subset of \numberCO\ of these galaxies. The sample is selected from fields covered by the {\it Herschel-}ATLAS survey that are also targeted by the MaNGA optical integral-field spectroscopic survey. The new JCMT observations combined with the multi-wavelength ancillary data will allow for the robust characterization of the properties of dust in the nearby Universe, and the benchmarking of scaling relations between dust, gas, and global galaxy properties. In this paper we give an overview of the survey objectives and details about the sample selection and JCMT observations, present a consistent 30 band UV-to-FIR photometric catalog with derived properties, and introduce the JINGLE Main Data Release (MDR). Science highlights include the non-linearity of the relation between 850$\mu$m luminosity and CO line luminosity ($\log L_{CO(2-1)}=1.372\log L_{850}$-1.376), and the serendipitous discovery of candidate $z>6$ galaxies.
\end{abstract}


\section{Introduction}
\label{intro}

The impact of large imaging and spectroscopic surveys on galaxy evolution studies has been substantial. Systematic observations of very large samples of galaxies at optical, ultraviolet (UV) and infrared (IR) wavelengths have, for example, allowed for precise measurements of stellar masses and star formation rates (SFRs) up to $z\approx3$. These measurements show how star-forming galaxies form a tight sequence in the SFR-\mstar\ plane whose shape is mostly redshift independent, but whose zero-point is shifted to ever higher SFRs as redshift increases \citep[e.g.][]{noeske07,rodighiero10,whitaker12}.

Although such large surveys at UV-to-IR wavelengths have been standard practice for decades, folding millimetre (mm) and radio spectral line observations into such multi-wavelength statistical studies is comparatively recent practice. New and improved instruments (e.g.\ multi-beam receivers on radio telescopes, and sensitive receivers and backends fitted to mm/sub-mm dishes) have recently sped up the process of accumulating these challenging observations, making it possible to add atomic and molecular gas masses to the list of physical properties measurable over large, representative galaxy samples \citep[e.g.][]{GASS1,COLDGASS1,tacconi13}. Such measurements have led to the understanding that galaxy evolution is driven to a large extent by the availability of cold gas in different galaxies at certain times and in particular environments, and, for example, can explain simply the redshift evolution of the main sequence \citep{saintonge13,sargent14}. Despite the technical challenges, further progress will only come from broadening the samples targeted for molecular gas studies, particularly focusing on galaxies with low stellar masses and objects beyond $z\sim2.5$. 

While measurements of the mass and properties of the cold interstellar medium (ISM) are typically obtained via molecular and atomic line spectroscopy, it has become increasingly common practice to use far-infrared (FIR)/sub-mm continuum observations of galaxies to derive total dust masses, from which total gas masses are in turn inferred via the gas-to-dust ratio \citep[e.g.\ ][]{israel97,leroy11,magdis11,eales2012,sandstrom13,scoville14,groves15}. This method has generated significant interest, as it allows for gas masses to be measured for very large samples much more quickly and cheaply than via direct CO (and H{\small I}) measurements. The technique is of particular interest for low-mass and/or high-redshift galaxies with low metallicities, where it is known that CO suffers from photodissociation effects. However, there are many unknowns in this method that must be investigated before it can be applied reliably at high redshifts. For example, a simple linear relation between gas-to-dust ratio and metallicity is currently assumed, while there are indications of a large scatter at fixed metallicity and a possible redshift evolution \citep{galametz11,saintonge13,remyruyer14,accurso17}. Furthermore, the dust masses are estimated assuming that dust in all galaxies has properties similar to those in the Milky Way, which are now known not to be universally applicable \citep[e.g.][]{gordon03,smith12,clayton15}. 

There is therefore a pressing need for a systematic survey of the dust properties in a variety of galaxies to benchmark scaling relations with gas content as well as stellar, chemical and structural properties. Such work will not only have profound implications for our understanding of gas and dust physics in nearby galaxies, but also for high-redshift work (either with the JCMT itself or with ALMA), where observers have to look beyond CO(1-0) spectroscopy to investigate the cold ISM.
Finally, even if the dust properties resemble those in the Milky Way, estimating the dust masses from a relatively small number of photometric measurements using a method based on fitting the temperature $T$ and opacity index $\beta$, as is commonly done, may suffer from systematic errors due to measurement errors, the assumed $T$-distributions being too simplistic (e.g. a single temperature, or only two distinct temperatures), and the $T$-dependence of $\beta$ itself, as demonstrated in laboratory measurements \citep{Mennella_98_Temperature,Boudet_05_Temperature,Coupeaud_11_Low,Mutschke_13_Far}.

In this paper, we introduce the {\it JCMT dust and gas In Nearby Galaxies Legacy Exploration}, JINGLE, a new survey for molecular gas and dust in nearby galaxies. The main objectives of the survey are to provide a comprehensive picture of dust properties across the local galaxy population and to benchmark scaling relations that can be used to compare dust and gas masses with global galaxy observables such as stellar mass (\mstar), star formation rate (SFR) and gas-phase metallicity. After describing the sample selection and survey strategy, we present the extensive multi-wavelength data products upon which JINGLE builds and the homogeneous catalog of measurements derived from them. We also report on highlights from the survey's early science papers. 

Throughout this paper, we refer to accompanying JINGLE papers: Smith et al. (hereafter \paperII) describes the SCUBA-2 observations and data reduction process, Xiao et al. (hereafter \paperIII) presents the data and first results based on the CO(2-1) observations, and De Looze et al. (hereafter \paperiv) presents the first JINGLE dust scaling relations.   

All rest-frame and derived quantities assume a \citet{chabrier03} IMF, and a cosmology with $H_0=70$\kms\ Mpc$^{-1}$, $\Omega_m=0.3$ and $\Omega_{\Lambda}=0.7$.

\section{Survey objectives and sample selection}
\label{objectives}

JINGLE is a SCUBA-2 survey at 850$\mu$m of \ntot\ galaxies, with about half of the galaxies also being observed in the CO $J$=2-1 line (hereafter, CO(2-1)) using the RxA3m instrument. The sample consists of \textit{Herschel-}detected galaxies probing the star formation main sequence above \mstar$=10^9$\msun\ as illustrated in Figure \ref{sampleSCUBA2}. Amongst several other data products, the JCMT observations importantly provide total, integrated molecular gas masses through the CO(2-1) line measurements as well as accurate dust masses from the modeling of the 850$\mu$m and other infrared photometric points. 

\subsection{Science goals}

JINGLE has been designed to achieve three broad scientific goals:

{\it 1. Star formation, star formation history and the total gas reservoir.} The CO(2-1) line is a relatively linear tracer of the bulk molecular gas, just like CO(1-0). Combining integrated CO spectra with two-dimensional data from the SDSS-IV MaNGA survey \citep{bundy15}, it is possible to study correlations between the total cold gas content and optically-resolved properties of galaxies. Of particular interest are how radial gradients in quantities such as metallicity, ionisation mechanism, stellar age, and star formation rate correlate with the total molecular gas content. The wide range of physical parameters across the JINGLE-MaNGA sample also will allow us to probe how deviations from the canonical Kennicutt-Schmidt law \citep{kennicutt1998,schmidt59} depend on spatially resolved quantities such as gradients in the ionised gas.

{\it 2. Dust mass and dust scaling relations.} In combination with far-infrared data from the {\it Herschel Space Observatory}, the 850$\mu$m fluxes from JINGLE can be turned into measurements of the global dust mass, temperature, and emissivity that are significantly more accurate than values obtained from \textit{Herschel} data alone \citep{sadavoy2013}. We use these measurements to test for possible correlations of dust properties, such as the dust-to-stellar mass ratio, with galaxy metallicity, mass, star formation rate, etc. The wide range of stellar masses, morphological types, and metallicities in JINGLE allows us to benchmark scaling relations, which can then be applied to samples of high-redshift galaxies, and to constrain chemical evolution models.

{\it 3. The relation between molecular gas and dust.} The combination of CO, HI and 850$\mu$m data allows us to investigate the correlation of the dust mass with atomic, molecular, and total gas mass, as well as to probe whether dust properties (emissivity, temperature, grain composition) correlate with the fraction of gas in the molecular phase. With reliable gas-to-dust mass ratios, JINGLE will establish whether and how this ratio varies with other galaxy properties such as stellar mass, metallicity, and star formation rate. Finally, these data are used to quantify how accurately the 250, 500, and 850$\mu$m luminosities can be used to infer gas masses in low-redshift galaxies \citep{eales2012,scoville14,groves15}. Understanding the nature and scatter of these correlations will provide a vital check on this technique, which is increasing in popularity at both low and high redshifts.

\subsection{Sample selection}

To achieve our science goals, we need to observe a statistically significant galaxy sample and obtain homogeneous data products with the JCMT, making use of both RxA3m and SCUBA-2. We also require the following ancillary multi-wavelength data products:
\begin{enumerate}
\item \textit{Herschel} photometry to combine with the JCMT $850$\,$\mu$m   fluxes to derive accurate dust masses, temperatures and emissivities;
\item optical integral field spectroscopy (IFS) to derive spatially-resolved (i.e.\ {\rm gradients}) stellar and ionised gas properties, including metallicities;
\item H{\small I} observations (at the minimum integrated measurements, but ideally resolved maps) to quantify atomic gas masses within the same physical region of the galaxies as the CO and dust measurements.
\end{enumerate}

We identified as the ideal fields the North Galactic Pole (NGP) region and three of the equatorial Galaxy And Mass Assembly (GAMA) fields (GAMA09, GAMA12 and GAMA15). These four fields are part of \textit{Herschel}-ATLAS \citep[\textit{H}-ATLAS;][]{Eales2010} and therefore have uniform, deep Herschel-SPIRE coverage, fulfilling our first requirement. The four fields are also all within the footprint of the MaNGA IFS survey, and the GAMA fields are further being covered by the Sydney-AAO Multi-object Integral-field spectrograph (SAMI), ensuring the availability of optical IFS information. Finally, all four fields are within the footprint of the  Arecibo Legacy Fast ALFA Survey (ALFALFA) survey, so integrated H{\small I} masses are already available for about half of the galaxies, and an ongoing Arecibo programme (PI: M. Smith) is targeting all other JINGLE targets. In addition, the NGP is a high priority field for the blind Medium Deep Survey to be conducted at Westerbork with the new APERTIF phased array feed. As for the three GAMA fields, they lie within the footprint of WALLABY, an all-(southern) sky HI survey with the Australian Square Kilometer Array Pathfinder (ASKAP). Both of these large scale blind HI surveys will give resolved H{\small I} maps on the timescale of a few years. 

\begin{figure}
    \centering
    \includegraphics[width=3.4in]{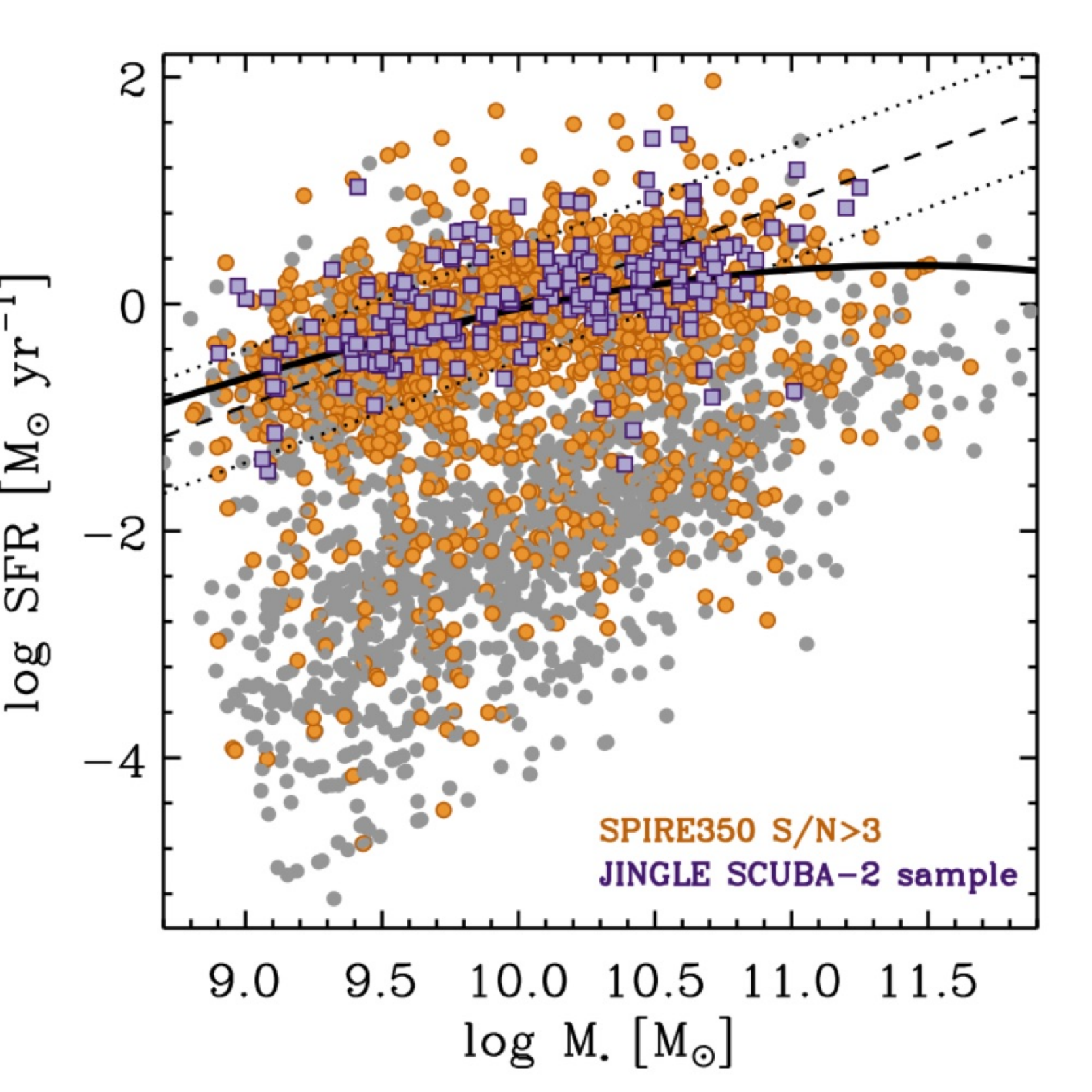}
    \caption{Distribution of the targeted and parent samples in the SFR-\mstar\ plane. All galaxies from the SDSS parent sample are shown, with those detected in the \textit{H}-ATLAS fields highlighted as coloured symbols. The purple squares represent the galaxies that are targeted with SCUBA-2 as part of JINGLE. We show the position of the star formation main sequence as determined by \citet{saintonge16} (solid line) and \citet{peng10} (dashed line, with 0.5 dex dispersion shown as dotted lines).}
    \label{sampleSCUBA2}
\end{figure}

We define as our parent sample for the selection of JINGLE targets all galaxies within our four fields that are part of the SDSS spectroscopic sample and have \mstar$>10^9$\msun\ and $0.01<z<0.05$. There are 2853 galaxies matching these selection criteria, out of which about half have been selected by MaNGA as possible targets. The distribution of the parent sample in the SFR-\mstar\ plane is shown in Figure~\ref{sampleSCUBA2}. 

Out of this parent sample, we consider for JCMT observations those galaxies with a detection at the 3$\sigma$ level at both 250 and 350$\mu$m in the \textit{H}-ATLAS survey. Given the depth of the \textit{H}-ATLAS SPIRE maps and the sensitivity of SCUBA-2, a galaxy with a far-infrared continuum detectable at 850$\mu$m before reaching the confusion limit would almost certainly be detected at both 250 and 350$\mu$m. The requirement for \textit{H}-ATLAS detections means that JINGLE targets are overwhelmingly selected from the blue star-forming galaxy population (Figure~\ref{sampleSCUBA2}). 

There are 284 galaxies in the parent sample that pass our \textit{Herschel} selection criterion at 250 and 350$\mu$m and also are predicted to be detectable with SCUBA-2 in less than 2 hours of integration. To have as uniform coverage as possible of the SFR-\mstar\ plane, we extracted 200 galaxies from this sub-sample in order to have a flat logarithmic stellar mass distribution. Since the mass distribution of the parent sample is well known, we can statistically correct for the flat stellar mass distribution a posteriori. This is a common procedure used by surveys such as GASS and MaNGA \citep[e.g.][]{GASS1}.  The final sample targeted for SCUBA-2 observation is presented in Fig. \ref{sampleSCUBA2}. The initial target selection was done using the stellar masses and SFRs released by \citet{chang15} and calculated with MAGPHYS \citep{MAGPHYS} using GALEX and SDSS photometry, while in Figs. \ref{sampleSCUBA2} and \ref{sampleRxA} (and throughout this paper), we make use of the new stellar masses derived specifically by the JINGLE team using MAGPHYS again,  but with our own 30-band multi-wavelength catalog (see Section \ref{data}). As will be shown in Fig. \ref{Mstarcomp}, the two sets of stellar masses follow each other linearly, with a systematic offset of 0.2 dex and a scatter of 0.15 dex. This explains why in the final JINGLE sample some galaxies have stellar masses just below $10^9$\msun. 

\begin{figure}
    \centering
    \includegraphics[width=3.4in]{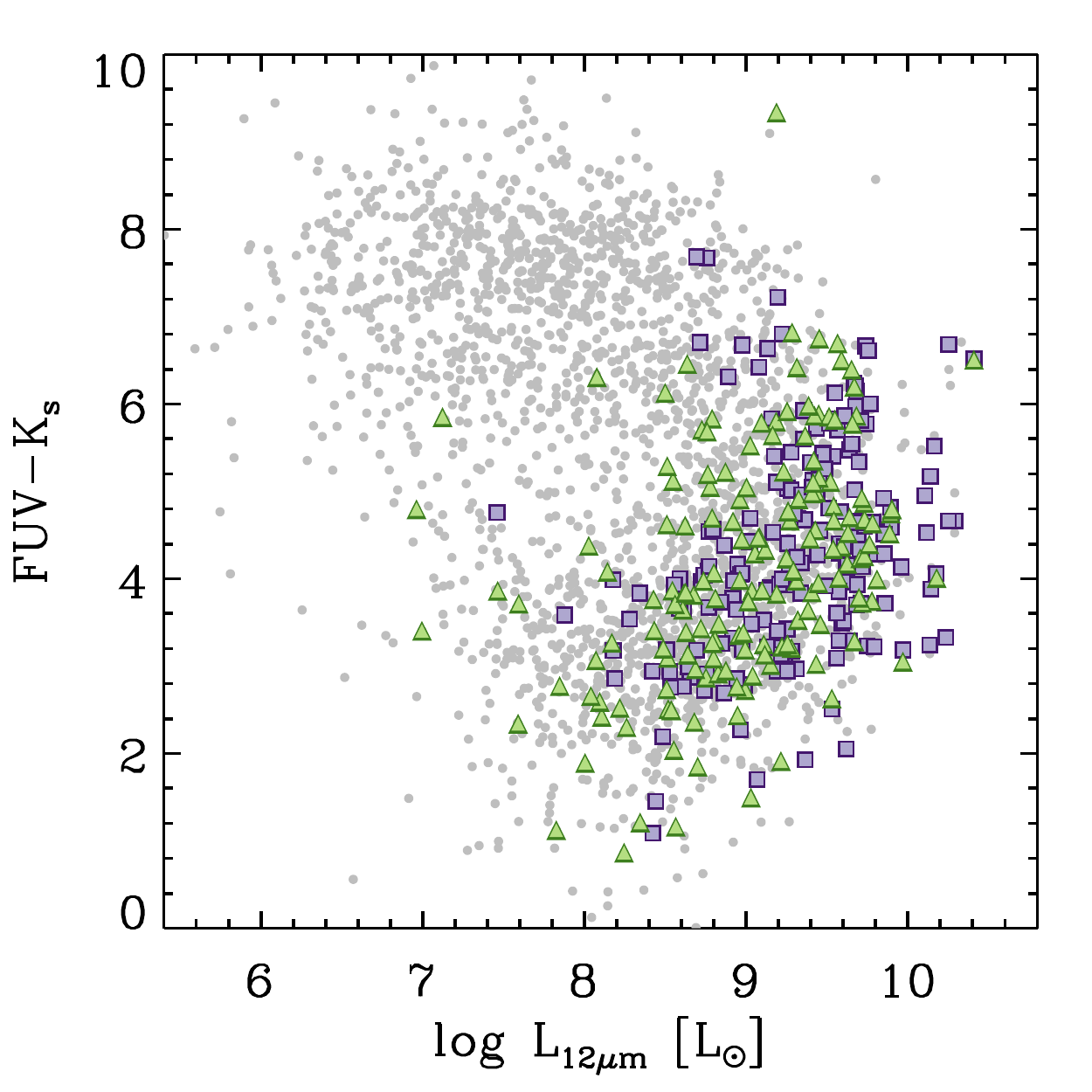}
    \caption{Colour-luminosity relation for the JINGLE sample (purple squares) and a matched control sample (green triangles), in comparison with the complete SDSS parent sample (gray circles).}
    \label{FUVKL12}
\end{figure}

\begin{figure}
    \centering
    \includegraphics[width=3.4in]{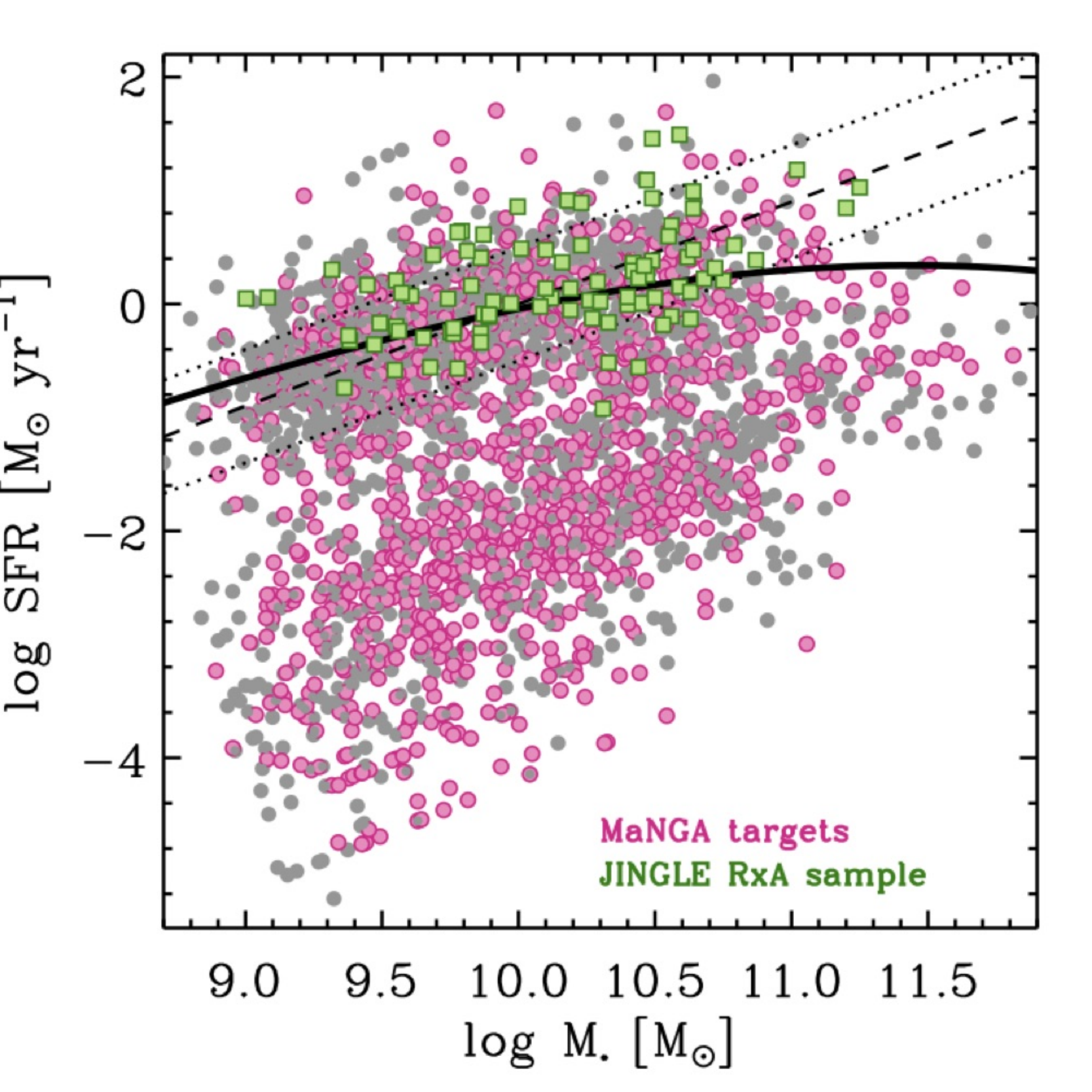}
    \caption{Distribution of the targeted and parent samples in the SFR-\mstar\ plane. The green squares show the subset of the SCUBA-2 sample (see Fig.\ref{sampleSCUBA2}) that are targeted for RxA3m observations as part of JINGLE. For comparison, the pink circles represent galaxies that are possible MaNGA targets. The position of the star formation main sequence as determined by \citet{saintonge16} (solid line) and \citet{peng10} (dashed line, with 0.5 dex dispersion shown as dotted lines) is also shown.}
    \label{sampleRxA}
\end{figure}

To test if the final JINGLE sample is biased towards particularly ISM-rich or dusty galaxies due to the selection criteria based on the \textit{Herschel}/SPIRE photometry, we construct a control sample extracted from the parent sample of 2853 galaxies which is only mass- and redshift-selected from SDSS. For each JINGLE galaxy, a control object is selected at random within 0.1 dex in \mstar\ and 0.2 dex in SFR. The process is repeated 150 times to produce a family of control samples. To assess whether the JINGLE galaxies are particularly dusty, in Figure~\ref{FUVKL12} we compare the distribution of the JINGLE sample and one randomly-chosen realisation of the control sample in the parameter space formed by WISE 12$\mu$m luminosity and FUV$-K_s$ colour. Colours such as FUV$-K_s$ or NUV$-r$ have been shown to correlate well with the HI gas-to-stellar mass ratio, and therefore describe to which extent galaxies are ISM-rich \citep{catinella13,devis17}. The Kolmogorov-Smirnov (KS) probability that the FUV$-K_s$ distribution of the JINGLE and control samples are extracted from the same underlying distribution is $0.50\pm0.23$; such a result indicates that the JINGLE sample is not biased towards particularly ISM-rich galaxies. 

However, as Figure \ref{FUVKL12} shows, there is a tendency for some JINGLE galaxies to have higher 12$\mu$m luminosities than their control objects. This is particularly evident for the redder population (FUV$-K_s>6$). Similarly, among the blue population, there is a tail of control galaxies with $L_{12\mu m}<10^8L_{\odot}$ which are mostly absent from the JINGLE sample, and vice versa. Indeed, the KS test, with a probability of $0.004\pm0.002$, confirms that the distributions of $L_{12\mu m}$ of the JINGLE and control samples are different, with the JINGLE objects shifted towards higher IR luminosities {(and therefore probably higher dust masses and/or stronger radiation fields)}. With on average normal FUV$-K_s$ colours but elevated 12$\mu$m luminosities, the JINGLE galaxies are possibly biased towards dust- or H$_2$-rich systems at fixed HI mass; this will have to be carefully corrected for in upcoming analyses of dust scaling relations. 

Out of the \ntot\ galaxies targeted with SCUBA-2, a subset of \numberCO\ objects predicted to be detectable in less than 14 hours of integration was selected to be observed with the heterodyne receiver RxA3m to obtain integrated CO(2-1) line fluxes. Galaxies that are part of the currently released MaNGA sample were given first priority for CO(2-1) observations, though all the galaxies selected for RxA3m observations are candidate MaNGA targets and likely to be part of future SDSS data releases. Figure \ref{sampleRxA} illustrates the position of the sample selected for RxA3m observations in the \ms\ plane.

\subsection{JCMT observations}

To plan for observations, predictions of 850\micron\ continuum and CO(2-1) line luminosities were made for all the galaxies in the JINGLE parent sample. Extensive details about these calculations as well as descriptions of the observing strategy and the data products associated with the SCUBA-2 and RxA3m components of the survey are presented in \paperII\ and \paperIII, respectively. A summary is presented here as an overview.  

\subsubsection{SCUBA-2}

The sub-millimeter continuum observations for JINGLE are obtained with SCUBA-2, the 10000 pixel bolometer camera operating at the JCMT \citep{holland13}. With two independent imaging arrays, SCUBA-2 can simultaneously map the sky at 450 and 850$\mu$m. Given the availability of 500$\mu$m fluxes from \textit{Herschel}, and the significantly lower atmospheric transmission at 450$\mu$m, the JINGLE survey is based on the requirement of detecting the continuum at 850$\mu$m. However, as we simultaneously observe at 450$\mu$m, for targets observed in better weather conditions there is the possibility of detecting higher-resolution 450$\mu$m dust continuum emission as well. 

To prepare for the observations, a single modified blackbody with $\beta=2$ was fitted to the \textit{Herschel} fluxes; this fit was extrapolated to estimate the 850$\mu$m flux. Given their angular sizes ($D_{25}=20-50$\arcsec) as well as the 13\arcsec\ beam of SCUBA-2 at 850$\mu$m, the JINGLE galaxies are marginally resolved in the maps. The integration time required for each galaxy to reach a 5$\sigma$ detection was determined through the SCUBA-2 exposure calculator, taking into account the galaxy's angular extent and assuming matched beam filtering and a range of weather conditions.  

Observations are conducted in Daisy mode, which provides uniform coverage over a central 4$^\prime$ region with significant coverage out to 12$^\prime$. The weather band (either grade 2, 3 or 4) was chosen so we would reach the required sensitivity in under two hours. To achieve this, JINGLE was awarded 255 hours of SCUBA-2 observing time, spread over weather bands 2, 3 and 4. The exact definition of the JCMT weather bands as a function of opacity at 225GHz and levels of precipitable water vapor (PWV) are available on the JCMT web pages\footnote{\url{http://www.eaobservatory.org/jcmt/observing/weather-bands/}}.

\subsubsection{RxA3m}
The CO(2-1) line fluxes were estimated from the specific star formation rate of each object using the depletion timescale and CO-to-H$_2$ conversion factor predicted by the 2-SFM formalism of \citet{sargent14}. {To validate these estimates, CO line fluxes were also extrapolated from the WISE 12$\mu$m luminosities using the calibration of \citet{2015ApJ...799...92J} and assuming a CO(2-1)/CO(1-0) line ratio of $r_{21}=0.7$ and a CO-to-H$_2$ conversion factor $\alpha_{\rm CO}=4.35$\msun (K \kms pc$^2)^{-1}$.} The integration times are set by the requirement to detect the predicted line flux at the 5$\sigma$ level over a spectral channel corresponding to 20\% of the expected (Tully-Fisher-inferred) line width. These integration times are calculated for weather bands 4 or 5 and the specific properties of the telescope and instrument. 

The survey was granted 525 hours of observing to complete the CO(2-1) observations, most of which is in band 5 to be used as a poor weather filler. At the frequency of the CO(2-1) line, the beam size is 20\arcsec, and given the angular size of the galaxies we observe in beam switching mode with a throw of 120\arcsec. The receiver bandwidth is 1000~MHz. Observations are monitored and reduced on a nightly bias.  If a secure line detection is reached before the estimated required sensitivity is reached, observations of that galaxy are stopped.  Otherwise, we continue observing the galaxy until the estimated sensitivity is reached. As is shown in \paperIII, given the necessary integration time, reliable detections of the CO(2-1) line can be achieved for the JINGLE galaxies under such weather conditions after smoothing the spectrum to 30~\kms.

\section{Ancillary data products and derived quantities}
\label{data}

JINGLE relies not only on its own JCMT data products but also on the availability of several ancillary data sets across the electromagnetic spectrum. In particular, the availability of the far-infrared photometry from \textit{Herschel} is key. Being a blind, wide-area survey of uniform depth with point source sensitivities of $7.4$, $9.4$ and $10.2$\,mJy ($1\sigma$ total noise) at $250$, $350$ and $500$\,$\mu$m \citep{valiante2016}, \textit{H}-ATLAS is perfectly suited to provide the deep, uniform FIR photometry required to achieve the science objectives of JINGLE. Maps of the GAMA fields are provided by \textit{H}-ATLAS data release 1 \citep{valiante2016} and the NGP field by data release 2 \citep{Smith2017}.
The other external survey which is an integral part of the JINGLE strategy is MaNGA as it will provide two-dimensional (i.e.\ spatially-resolved) measurements of the stellar mass surface density, kinematics and chemical element abundance ratio for a significant fraction of the JINGLE galaxies for which CO(2-1) observations are conducted. However, as both JINGLE and MaNGA are ongoing surveys, the number of galaxies with both JCMT data products in the JINGLE Main Data Release and MaNGA data products in SDSS DR14 \citep{SDSSDR14} is low, and joint analyses will therefore be the topic of future papers.  

Here however, we make use of the abundant photometry available through \textit{H}-ATLAS as well as a range of all-sky legacy surveys to construct a uniform multi-wavelength flux catalog for the JINGLE objects and derive important physical quantities such as stellar masses and star formation rates. 

\begin{table*}
\begin{center}
\footnotesize
\caption{Details of each band for which we produced CAAPR photometry. For FUV--$K_{S}$ bands, we refer to each band by its listed `Band Name'; otherwise we refer to bands by wavelength. The `Photometry Present' column gives the number of galaxies in each band for which we present photometry (not counting photometry excluded due to image artefacts or insufficient sky coverage). References for calibration uncertainties and data archives are provided in the table footnotes.}
\label{CAAPR_Table}
\begin{tabular}{lrlSSSSll}
\toprule \toprule
\multicolumn{1}{c}{Facility} &
\multicolumn{1}{c}{Effective} &
\multicolumn{1}{c}{Band} &
\multicolumn{1}{c}{Photometry} &
\multicolumn{1}{c}{Pixel} &
\multicolumn{1}{c}{Resolution} &
\multicolumn{2}{c}{Calibration} &
\multicolumn{1}{c}{Data} \\
\multicolumn{1}{c}{} &
\multicolumn{1}{c}{Wavelength} &
\multicolumn{1}{c}{Name} &
\multicolumn{1}{c}{Present} &
\multicolumn{1}{c}{Width} &
\multicolumn{1}{c}{FWHM} &
\multicolumn{2}{c}{Uncertainty} &
\multicolumn{1}{c}{Archive} \\
\cmidrule(lr){5-5}
\cmidrule(lr){6-6}
\cmidrule(lr){7-8}
\multicolumn{1}{c}{} &
\multicolumn{1}{c}{} &
\multicolumn{1}{c}{} &
\multicolumn{1}{c}{} &
\multicolumn{1}{c}{(\arcsec)} &
\multicolumn{1}{c}{(\arcsec)} &
\multicolumn{2}{c}{(\%)} &
\multicolumn{1}{c}{} \\
\midrule
GALEX & 153\,nm & FUV & 183 & 2.5 & 4.3 & 4.5 & \rdelim\}{2}{13pt}[$a$] & \rdelim\}{2}{13pt}[$b$] \\
GALEX & 227\,nm & NUV & 185 & 2.5 & 5.3 & 2.7 & & \\
SDSS & 353\,nm & {\it u} & 193 & 0.4 & 1.3 & 1.3 & \rdelim\}{5}{13pt}[$c$] & \rdelim\}{5}{13pt}[$d$] \\
SDSS & 475\,nm & {\it g} & 192 & 0.4 & 1.3 & 0.8 & & \\
SDSS & 622\,nm & {\it r} & 193 & 0.4 & 1.3 & 0.8 & & \\
SDSS & 763\,nm & {\it i} & 192 & 0.4 & 1.3 & 0.7 & & \\
SDSS &  905\,nm& {\it z} & 193 & 0.4 & 1.3 & 0.8 & & \\
VISTA & 877\,nm & {\it Z} & 45 & 0.4 & 0.8 & 2.7 & \rdelim\}{5}{13pt}[$e$] & \rdelim\}{5}{13pt}[$f$] \\
VISTA & 1.02\,\micron\ & {\it Y} & 44 & 0.4 & 0.8 & 2.7 & & \\
VISTA & 1.25\,\micron\ & {\it J} & 12 & 0.4 & 0.8 & 2.7 & & \\
VISTA & 1.65\,\micron\ & {\it H} & 45 & 0.4 & 0.8 & 2.7 & & \\
VISTA & 2.15\,\micron\ & {\it K$_{S}$} & 47 & 0.4 & 2.0 & 2.7 & & \\
2MASS & 1.24\,\micron\ & {\it J} & 192 & 1 & 2.0 & 1.7 & \rdelim\}{3}{13pt}[$g$] & \rdelim\}{7}{13pt}[$h$] \\
2MASS & 1.66\,\micron\ & {\it H} & 191 & 1 & 2.0 & 1.9 & & \\
2MASS & 2.16\,\micron\ & {\it K$_{S}$} & 192 & 1 & 2.0 & 1.9 & & \\
WISE & 3.4\,\micron\ & (W1) & 182 & 1.375 & 6.1 & 2.9 & \rdelim\}{4}{13pt}[$i$] & \\
WISE & 4.6\,\micron\ & (W2) & 183 & 1.375 & 6.4 & 3.4 & & \\
WISE & 12\,\micron\ & (W3) & 193 & 1.375 & 6.5 & 4.6 & & \\
WISE & 22\,\micron\ & (W4) & 193 & 1.375 & 12 & 5.6 & & \\
{\it Spitzer} & 4.5\,\micron\ & (IRAC-2) & 28 & 0.6 & 1.72 & 3 & \rdelim\}{3}{13pt}[$j$] & \rdelim\}{6}{13pt}[$k$] \\
{\it Spitzer} & 5.8\,\micron\ & (IRAC-3) & 17 & 0.6 & 1.88 & 3 & & \\
{\it Spitzer} & 8.0\,\micron\ & (IRAC-4) & 16 & 0.6 & 1.98 & 3 & & \\
{\it Spitzer} & 24\,\micron\ & (MIPS-1) & 25 & 2.45 & 6 & 5 & \rdelim\}{3}{13pt}[$l$] & \\
{\it Spitzer} & 70\,\micron\ & (MIPS-2) & 18 & 4 & 18 &10 & & \\
{\it Spitzer} & 160\,\micron\ & (MIPS-3) & 18 & 8 & 38 & 12 & & \\
{\it Herschel} & 100\,\micron\ & (PACS-Green) & 190 & 3 & 11 & 7 & \rdelim\}{2}{13pt}[$m$] & \rdelim\}{5}{13pt}[$n$] \\
{\it Herschel} & 160\,\micron\ & (PACS-Red) & 190 & 4 & 14 & 7 & & \\
{\it Herschel} & 250\,\micron\ & (SPIRE-PSW) & 193 & 6 & 18 & 5.5 & \rdelim\}{3}{13pt}[$o$] & \\
{\it Herschel} & 350\,\micron\ & (SPIRE-PMW) & 193 & 8 & 25 & 5.5 & & \\
{\it Herschel} & 500\,\micron\ & (SPIRE-PLW) & 193 & 12 & 36 & 5.5 & & \\
\bottomrule
\end{tabular}
\end{center}
\footnotesize
\justify
\begin{multicols}{2}{
$^{a}$ \citet{Morrissey2007B} \\
$^{b}$ Mikulski Archive for Space Telescopes (MAST): \url{http://galex.stsci.edu/GR6/} \\
$^{c}$ SDSS DR12 Data Release Supplement: \url{https://www.sdss3.org/dr12/scope.php} \\
$^{d}$ SDSS DR12 Science Archive Server: \url{https://dr12.sdss.org/home} \\
$^{e}$ VISTA Instrument Description: \url{https://www.eso.org/sci/facilities/paranal/instruments/vircam/inst.html} \\
$^{f}$ VISTA Science Archive: \url{http://vsa.roe.ac.uk/} \\
$^{g}$ \citet{Cohen2003E} \\
$^{h}$ NASA/IPAC Infrared Science Archive (IRSA): \url{http://irsa.ipac.caltech.edu} \\
$^{i}$ WISE All-Sky Release Explanatory Supplement: \url{http://wise2.ipac.caltech.edu/docs/release/allsky/expsup/sec4_4h.html} \\
$^{j}$ IRAC Instrument Handbook: \url{https://irsa.ipac.caltech.edu/data/SPITZER/docs/irac/iracinstrumenthandbook/17/\#_Toc410728305} \\
$^{k}$ {\it Spitzer} Heritage Archive (SHA): \url{http://sha.ipac.caltech.edu/applications/Spitzer/SHA/} \\
$^{l}$ MIPS Instrument Handbook: \url{https://irsa.ipac.caltech.edu/data/SPITZER/docs/mips/mipsinstrumenthandbook/42/\#_Toc288032317} \\
$^{m}$ PACS Instrument \& Calibration Wiki:  \url{http://herschel.esac.esa.int/twiki/bin/view/Public/PacsCalibrationWeb} \\
$^{n}$ {\it Herschel}-ATLAS: \url{http://www.h-atlas.org/public-data/download} \\
$^{o}$ SPIRE Instrument \& Calibration Wiki: \url{http://herschel.esac.esa.int/twiki/bin/view/Public/SpireCalibrationWeb} \\}
\end{multicols}
\end{table*}

\subsection{Multiwavelength photometry}
\label{caapr}

A key feature of JINGLE is the uniformity of the dust and gas measurements being gathered, since all the observations are conducted with the same instruments and to consistent depths. To best exploit this feature, it is essential that all physical parameters (stellar masses, SFRs, metallicities, etc.) are derived in a consistent manner. To this end, we have produced an extensive 30-band multi-wavelength photometric catalog. This catalog makes use of data from 7 UV--submm facilities: the GALaxy Evolution eXplorer (GALEX; \citealp{Morrissey2007B}), the Sloan Digital Sky Survey (SDSS; \citealp{York2000B,Eisenstein2011B}) the 2 Micron All-Sky Survey (2MASS; \citealp{Skrutskie2006A}), the Visible and Infrared Survey Telescope for Astronomy (VISTA; \citealp{Sutherland2015B}), the Wide-field Infrared Survey Explorer (WISE; \citealp{Wright2010F}), the {\it Spitzer} Space Telescope, \citep{Werner2004B}, and {\it Herschel}. Table~\ref{CAAPR_Table} summarises important parameters for all these bands . All imagery was obtained from the official archives of each facility (except for the {\it Herschel} data, which is provided by {\it Herschel}-ATLAS); the data acquisition process was identical to that used in \citet{clark2017}.

The aperture-matched photometry was performed using the Comprehensive Adjustable Aperture Photometry Routine (CAAPR\footnote{\url{https://github.com/Stargrazer82301/CAAPR}}) pipeline, described in detail in \citet{clark2017}; CAAPR is a development of the photometry pipeline used in \citet{clark15} and \citet{devis17}. 

Before being able to perform photometry, contamination from foreground stars in the UV--MIR bands was minimised using the star-removal code contained in the Python Toolkit for SKIRT \citep[PTS;][]{camps15}. CAAPR removes any large-scale background structure (arising from cirrus, instrumental effects, etc.) by attempting to fit a 5th-order, 2-dimensional polynomial to the map (with the target galaxy and other bright sources masked). If the fitted polynomial is found to be significantly different from a flat sky, then CAAPR subtracts the polynomial from the map before proceeding with the rest of the photometry.

To make fluxes directly comparable across bands, aperture-matched photometry is performed. For each galaxy, elliptical apertures were fit to the source in each band; these apertures were then compared and combined to produce a `master' elliptical aperture that would enclose the source in every band. When performing this comparison, the sizes of the apertures were corrected to adjust for the PSF in each band by subtracting in quadrature the PSF FWHM major and minor axes of the aperture ellipse (effectively deconvolving them). Likewise, when performing the actual photometry using the master aperture, CAAPR convolves the aperture with each band's beam by adding in quadrature the major and minor axes of the aperture ellipse to the PSF FWHM. 

An annulus (with inner and outer major axes 1.25 and 1.5 times the major axis of the source aperture, and the same position angle and axial ratio as the source aperture) was used to find the local background, which was estimated using an iteratively sigma-clipped median. For maps with pixel width \textgreater\,5$^{\prime\prime}$\ (i.e. the SPIRE bands) the flux inside apertures is measured with consideration for partial pixels. CAAPR determines the aperture noise associated with each flux value by randomly placing copies of the photometric apertures on the map around the source. All random apertures were positioned so as to avoid overlap with the actual source aperture as well as to avoid significant overlap with other random apertures. Although random, the apertures were biased towards being placed in regions of the map closer to the target source, according to a Gaussian distribution centered on the source coordinates. Fluxes in the random apertures were measured in the same way as for the source itself (i.e. including background annulus). The iteratively sigma-clipped standard deviation of these sky fluxes was taken as the aperture noise; this method thus incorporates instrumental noise and confusion noise. 

For bands with beam FWHM\,\textgreater\,5\,$^{\prime\prime}$, an aperture correction was applied to account for the fraction of the source flux spread outside the source aperture (and into the background annulus) by the PSF. Most instrument handbooks only provide such corrections for point sources, as corrections for extended sources (such as the JINGLE galaxies) require a model for the underlying unconvolved flux distribution. CAAPR assumes that each target galaxy, as observed in a given band, can be approximated as a 2-dimensional S\'ersic distribution convolved with the band's PSF. Therefore CAAPR fits a 2-dimensional PSF-convolved-S\'ersic model to the map, and uses the (unconvolved) S\'ersic distribution of the best-fit model to estimate the factor by which the measured flux is altered by the PSF. This factor was used to correct the measured flux accordingly. When performing these convolutions we use the circularised PSF kernels\footnote{\url{http://www.astro.princeton.edu/~ganiano/Kernels.html}} of \citet{aniano11} for all bands (for consistency). {The median value of the aperture correction in any given wave band is a function of the size of the PSF, and ranges for example from 1.01 for GALEX NUV (PSF FWHM: 5.3\arcsec), to 1.17 for PACS 100$\mu$m (FWHM 11\arcsec)  and 1.47 for SPIRE 500$\mu$m (FWHM 36\arcsec)}. No attempt to apply aperture corrections was made for sources with SNR\,\textless\,3, as the results of the fit were likely to be spurious.

Fluxes at wavelengths shorter than 10\,\micron\ were corrected for Galactic extinction according to the prescription of \citet{schlafly11}, using the IRSA Galactic Dust Reddening and Extinction Service\footnote{\url{https://irsa.ipac.caltech.edu/applications/DUST/}}.

The imagery and photometry was visually inspected and fluxes corrupted by image artefacts, etc, were removed. \citet{clark2017} provides detailed validation of CAAPR's photometric methodology for all bands, with the exception of the VISTA data, which is an extra addition for the JINGLE catalog. VISTA provides far superior NIR photometry where available (i.e., in the GAMA fields) than 2MASS, with dramatically smaller uncertainties (thanks to modern instrumentation, and the minimal sky noise at the VISTA Paranal site). For the sources where VISTA and 2MASS overlap, they have median flux ratios in  $J$, $H$, and $K_{S}$ band of 0.999, 0.970, and 1.007 respectively (for $>5\sigma$ fluxes only); these typical offsets are far smaller than the instruments' calibration uncertainties, and rule out any systematic deviations between the datasets. 

An example of this photometry, consistently derived from GALEX FUV to {\it Herschel} 500\micron, is shown for a typical JINGLE galaxy in Fig. \ref{SEDexample}, with the spectral energy distributions (SEDs) for the entire JINGLE sample compiled in Appendix \ref{SEDsection}.

\begin{figure*}
    \centering
    \includegraphics[width=7.0in]{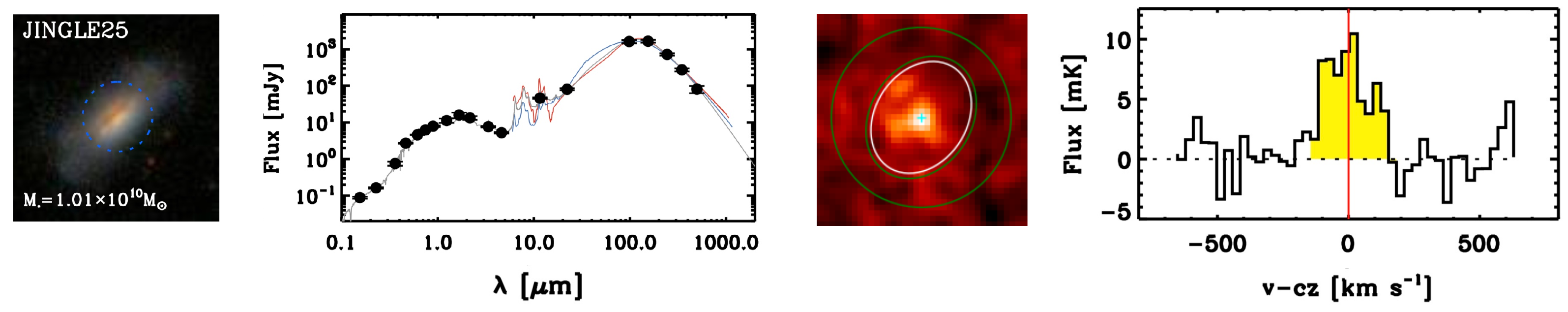}
    \caption{Example of the data products available as part of the JINGLE multiwavelength dataset and the Main Data Release (MDR) catalog. {\it Left:} 1\arcmin$\times$1\arcmin\ SDSS image centered on the position of the galaxy JINGLE25 (SDSSJ130636.39+275222.6). {\it Center left:} UV-to-FIR spectral energy distribution of this galaxy from the CAAPR photometric catalog. The best-fitting MAGPHYS model is shown as the gray line, as are the fits to the data points with $\lambda>30$\micron\ using the templates of \citet{ce01} renormalised following \citet{hwang10} (CE01; red line), and the hybrid AGN+SF templates of \citet{mullaney11} as implemented in \citet{hwang13} (JRM; blue line). {\it Center right:} JCMT SCUBA-2 continuum image of JINGLE25 at 850$\mu$m, 2.5\arcmin$\times$2.5\arcmin. The white ellipse shows the shape and position of the aperture used to measure the flux, while the region between the two green ellipses is used to determine the background. {\it Right:} JCMT RxA3m spectrum of this same galaxy, centered on the frequency of the CO(2-1) line.}
    \label{SEDexample}
\end{figure*}

\subsection{Star formation rates}
\label{SFRsection}

The CAAPR photometry was used to compute SFRs using a range of techniques, taking advantage of the broad wavelength coverage and the consistent photometry. Given the strong FIR/submm emphasis of JINGLE, we focus on SFR indicators that make use of these long wavelength data, although several tracers that involve only optical or UV data have also been calibrated and compared as part of the extensive analysis of \citet{davies16}. 
The techniques used fall in two categories: those which combine measurements of the unobscured and obscured SFRs from UV and IR photometry, and those which use the full multi-wavelength catalog and physical models taking energy balance into consideration. As an additional comparison, we also retrieved SFRs from the MPA/JHU catalog\footnote{\url{http://wwwmpa.mpa-garching.mpg.de/SDSS/}} for the JINGLE galaxies. These SFRs are derived from emission line fluxes within the SDSS fibers and aperture corrections based on the optical photometric colours \citep{brinchmann04}, and therefore represent a third, independent category of SFR estimates. 

\begin{figure*}
    \centering
    \includegraphics[width=6in]{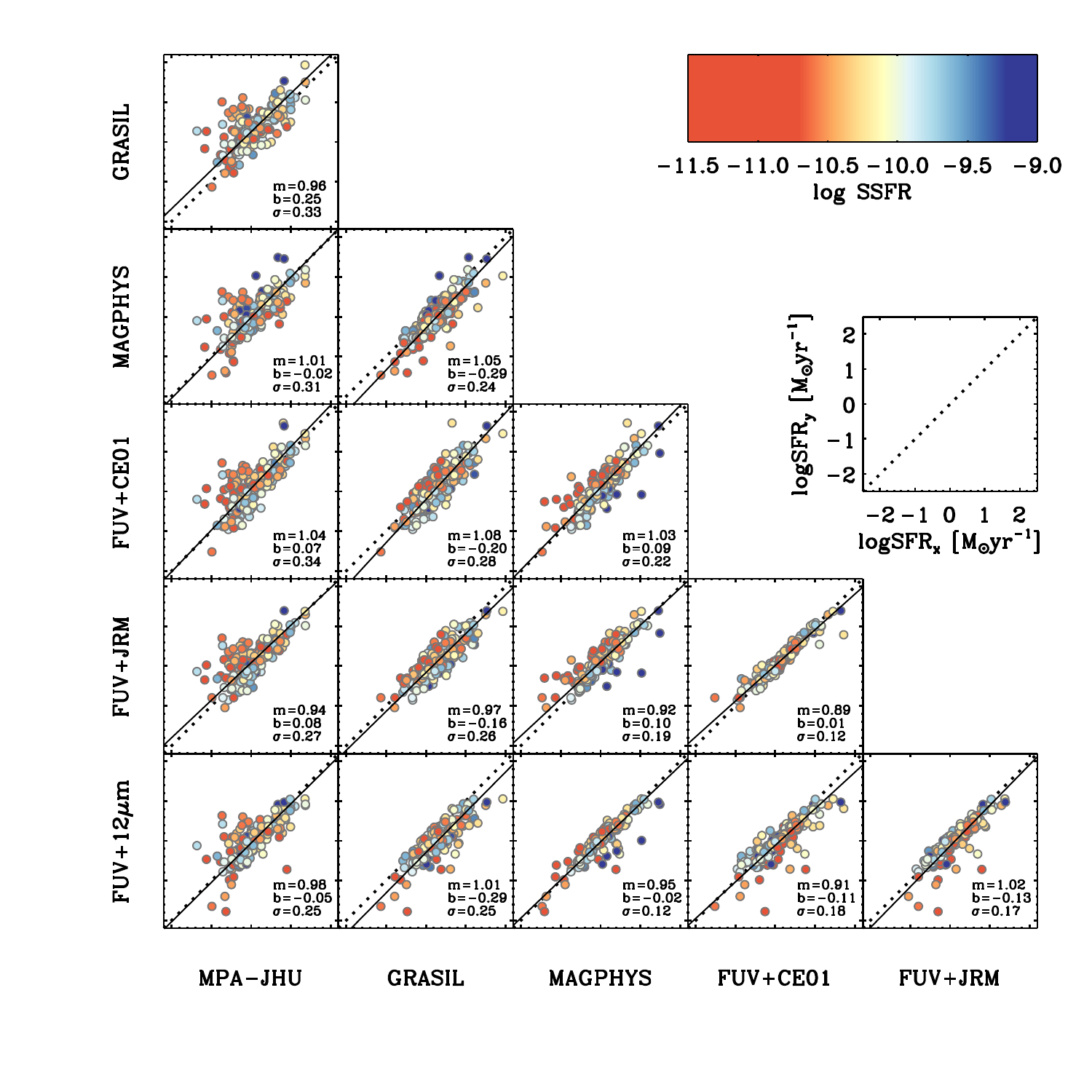}
    \caption{Comparison between the different SFR estimates calculated for the JINGLE sample using the CAAPR photometry, and those from SDSS photometry as retrieved from the MPA/JHU catalog. See \S\ref{SFRsection} for a description of the different SFR models. Dotted lines show a 1:1 relation and solid lines show a linear fit to the data, with the best fitting slope ($m$), intercept ($b$) and scatter ($\sigma$) given in each panel. Individual galaxies are colour-coded by sSFR.}
    \label{SFRcomp}
\end{figure*}

We briefly explain the different methods implemented with the CAAPR photometry. These are all compared against each other, and with the SDSS values, in Figure \ref{SFRcomp}. We have calculated three different flavours of SFRs within the first category; they all work by estimating separately \sfruv\ and \sfrir\ and taking the sum of the two as the total SFR:
\begin{itemize}
\item FUV+CE01: \sfruv\ is obtained directly from the {\it GALEX} FUV luminosity using the calibration presented in \citet{kennicuttevans12} and \sfrir\ is obtained by fitting the templates of \citet{ce01} for star-forming galaxies to all photometric data points with $\lambda>30$\micron, allowing renormalisation of the templates following \citet{hwang10}. 
\item FUV+JRM: \sfruv\ as above, but \sfrir\ is obtained using the templates of \citet{mullaney11} to all photometric points with $\lambda>20$\micron\ as done in \citet{hwang13}. The main difference with CE01 is that these templates take into account a possible AGN contribution to the FIR fluxes. 
\item {FUV+12\micron: \sfruv\ is here calculated from the {\it GALEX} FUV flux using the calibration of \citet{schiminovich07}, while \sfrir\ is derived from the WISE 12\micron\ fluxes using the calibration of \citet{jarrett13} and including a correction for stellar contamination using the WISE 3.4\micron\ fluxes following \citet{ciesla14}. A description and analysis of this method is presented in \citet{janowiecki17}. Unlike the others above, this SFR estimate is free of assumptions on the shape of the IR spectral energy distribution, although the related downside is that it does not consider possible systematic variations of the IR SED across the galaxy population \citep[e.g.][]{nordon12,boquien16}. }
\end{itemize}

The second category of SFRs are estimates obtained with two codes which use simple stellar population templates and models for the dusty ISM to reproduce the full SEDs of galaxies. First, {\sc magphys} \citep{MAGPHYS} was used to derive SFRs. {\sc magphys} is a panchromatic SED fitting tool capable of modelling the stellar and dust emission in galaxies under the assumption of a dust energy balance (i.e., the stellar energy that has been absorbed by dust is assumed to be re-emitted in the infrared). The stellar emission is modelled using \citet{bc03} stellar population models, assuming a \citet{chabrier03} IMF. The evolution of different stellar populations is calculated based on an analytic prescription of a galaxy's star formation history (SFH) represented as an exponentially declining star formation rate with some randomly imposed bursts. Dust attenuation of these stars is modelled using the two-phase model of \citet{charlot00}, and differentiates between young stars ($<$10$^{7}$\,yr) in dense molecular clouds attenuated by dust in their birth clouds and the ambient ISM dust, and older stars which only experience attenuation from the ambient ISM dust. The dust emission consists of the combined contribution of dust in birth clouds and in the ambient ISM. The dust emission in birth clouds is modelled using pre-defined templates for the emission of PAHs and transiently heated hot grains, and a modified blackbody (MBB) function with dust emissivity index $\beta$=1.5 and dust temperature T$_{\text{d}}$ between 30\,K and 70\,K for the emission of warm dust grains. An additional cold dust component (with $\beta$=2 and T$_{\text{d}}$ between 10\,K and 30\,K) is considered to model the dust emission from the ambient ISM. The latter temperature ranges correspond to the extended {\sc magphys} libraries from \citet{viaene14}. The dust masses in {\sc magphys} have been derived based on a dust mass absorption coefficient $\kappa_{\text{abs}}$(850\,$\mu$m)~=~0.77 cm$^{2}$ g$^{-1}$ \citep{dunne00}. Based on a Bayesian fitting algorithm, the best fitting stellar$+$dust emission model is derived from the libraries of 25,000 stellar population models and 50,000 dust emission spectra. Since the templates for the optical part of the
SED fitting come from \citet{bc03}, the model should not be biased against passive galaxies, an advantage over some of the methods described above. The best-fitting models can be seen for all the JINGLE galaxies in Appendix A.

In addition, we applied {\sc grasil} \citep{GRASIL} to all the SEDs; this code also includes templates suitable for a broad range of galaxies as well as the effects of dust. The templates used are from \citet{iglesias07} and the fitting technique is described in more detail in \citet{michalowski10}. In brief, {\sc grasil} is an SED fitting tool including radiative transfer that is coupled to a chemical evolution code \citep[CHE$\_$EVO,][]{silva99} and models the SFH of galaxies following a Kennicutt-Schmidt-type law \citep{schmidt59,kennicutt98}: SFR(t) = $\nu$M$_{\text{g}}$(t)$^{k}$ where $k$=1 and $\nu$ is a free parameter. The star formation rate is thus regulated by the gas mass which depends on the infall of primordial gas with a rate that is proportional to $\exp$(-$t$/$\tau_{inf}$), where the timescale $\tau_{inf}$ is a free parameter ranging between 0.1 and 21.6 Gyr. To mimic a recent burst of star formation, an extra star formation law with a declining timescale of 50\,Myr has been added to the SFH. To model the dust emission, {\sc grasil} considers three components: star-forming giant molecular clouds (GMCs), stars that have already emerged from their birth clouds, and diffuse gas. The timescale for stars to escape from molecular clouds, $t_{\text{esc}}$, is a free parameter of the model (varied from 1 to 4$\times$10$^{7}$ yr). Galaxies are modelled to have an age of 13\,Gyr and an exponential disk geometry with scale length of 4\,kpc and scale height of 0.4\,kpc with a range of inclinations (15, 45 and 75$^{\circ}$). The dust-to-gas ratio is assumed to be proportional to the metallicity. The dust emission from each model galaxy geometry is then calculated with a radiative transfer code. The dust masses from {\sc grasil} have been derived based on average dust opacities in the \citet{laor93} dust model with $\kappa_{\text{abs}}$(250\,$\mu$m)~=~6.4 cm$^{2}$ g$^{-1}$.

As shown in Fig. \ref{SFRcomp}, there is generally good agreement between all possible pairs of SFR indicators with scatter in the range of 0.1-0.3 dex. As expected, the tightest correlations are seen between indicators that are closely related, such as FUV+CE01 and FUV+JRM. The largest scatter is observed in the comparisons that involve the MPA/JHU spectral values. For these nearby galaxies, aperture corrections have to be applied to these spectral measurements as the SDSS fibers cover 3\arcsec\ while the optical diameters of our galaxies are typically 20-60\arcsec. These aperture corrections  could explain some of the scatter compared with methods that use the integrated flux from the galaxies. Most pairs of indicators have best-fitting slopes that are linear and with no systematic offsets, with the exception of the {\sc grasil} SFRs which are systematically larger than the other indicators by 0.1-0.2 dex. 

A priori, the {\sc magphys} SFRs would be expected to be best across the JINGLE sample, which includes both star-forming galaxies and massive galaxies below the main sequence. Indeed, the comparison between {\sc magphys} and FUV+CE01 and FUV+JRM shows how galaxies with the highest and lowest specific star formation rates (sSFRs) scatter the most from the 1:1 relation. In comparison, the agreement between the {\sc magphys} and the FUV+12\micron\ values is better with a scatter of only 0.12 dex. {The systematic offset between the {\sc magphys} and FUV+12\micron\ values for the galaxies with the highest SSFRs is likely due to the latter not accounting for systematic variations in the shape of the IR spectral energy distribution as galaxies move away from the main sequence.} From all these comparisons, we adopt the {\sc magphys} and FUV+12\micron\ values as the main JINGLE SFR estimates; as they are mostly independent from each other they will allow us to test that any result is not dependent on the particular SFR measurement used. All the other SFRs we have computed and compiled are however made available as part of the data release, to aid with comparison between JINGLE and other studies.

\subsection{Stellar masses}
\label{mass_section}

\begin{figure*}
    \centering
    \includegraphics[width=6in]{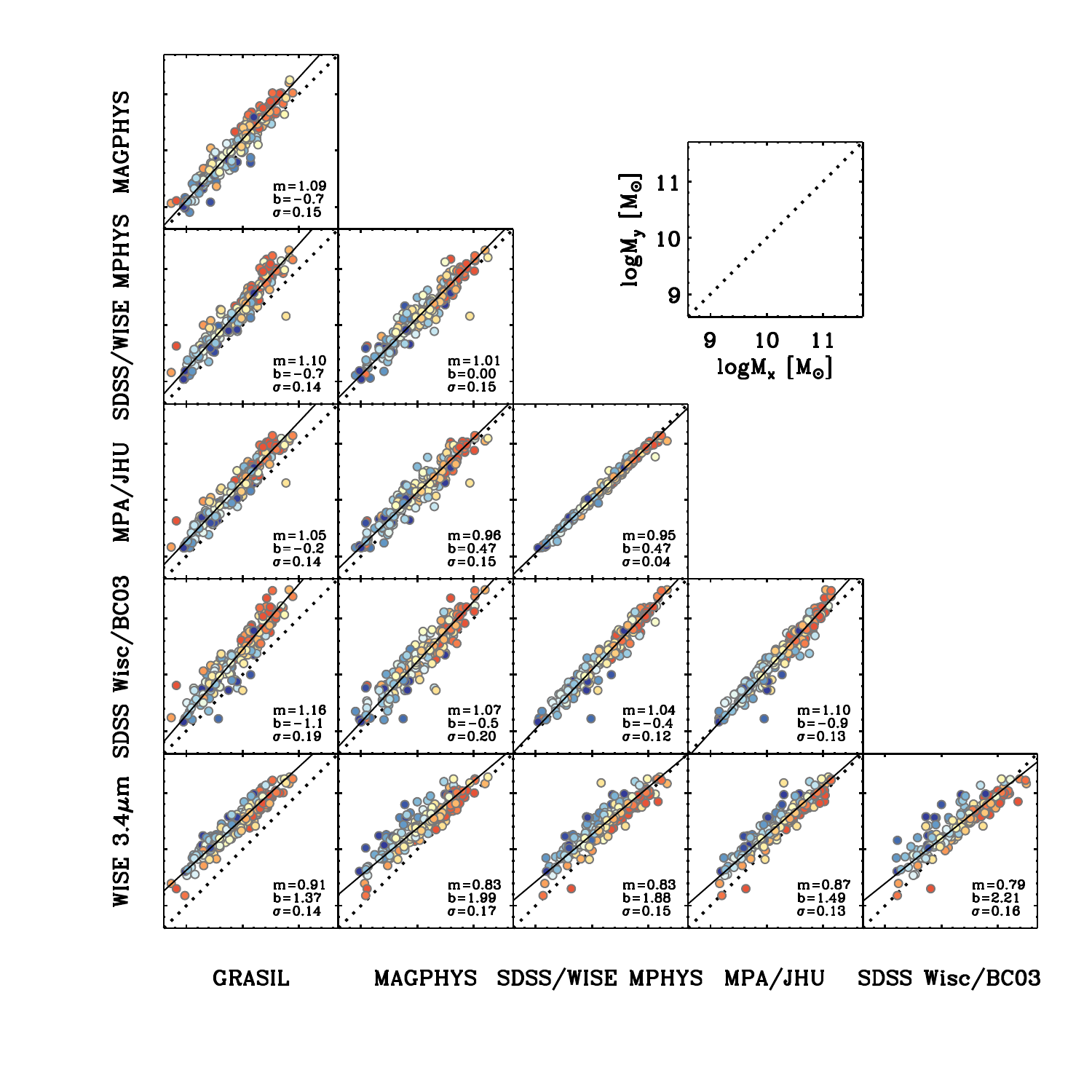}
    \caption{Comparison between the different stellar mass estimates calculated for the JINGLE sample using the CAAPR photometry, and those from SDSS photometry as retrieved from the DR10 database. See \S\ref{mass_section} for a description of the different SFR models. Dotted lines show a 1:1 relation and solid lines show a linear fit to the data, with the best fitting slope ($m$), intercept ($b$) and scatter ($\sigma$) given in each panel. Points are colour-coded according to specific star formation rate as in Fig.~\ref{SFRcomp}.}
    \label{Mstarcomp}
\end{figure*}

We have calculated stellar masses for all JINGLE galaxies from the CAAPR photometry as part of the MAGPHYS and GRASIL fitting. Additionally, the CAAPR-measured WISE 3.4$\mu$m luminosities are used to estimate \mstar\ by assuming a constant mass-to-light ratio of 0.47 \citep{mcgaugh14}. In Figure \ref{Mstarcomp}, these stellar masses are compared with three alternative estimates:
\begin{itemize}
\item SDSS/WISE MPHYS: from \citet{chang15}, an independent determination of \mstar\ using MAGPHYS, making use of SDSS and WISE photometry
\item MPA/JHU: from the MPA-JHU catalog\footnote{\url{http://wwwmpa.mpa-garching.mpg.de/SDSS/DR7/}}, these \mstar\ values are based on the SDSS photometry and calculated following \citet{salim07}
\item SDSS Wisc/BC03: these \mstar\ values are retrieved from the SDSS DR10 database, and have been calculated using the PCA-based method of \citet{chen12} and stellar population models from \citet{bc03}
\end{itemize}

The scatter between pairs of different \mstar\ measurements is in the range of 0.1 to 0.3 dex. The scatter is largest and the relations farthest from linear when comparing any mass estimate with the one calculated from the WISE 3.4$\mu$m luminosities, suggesting that the assumption of a constant mass-to-light ratio is not appropriate across the JINGLE sample, {or that dust is a contributor to the 3.4$\mu$m luminosities \citep{meidt14}}. In the rest of this paper we adopt the values of \mstar\ from MAGPHYS and the CAAPR photometry, but all other estimates are also made available as part of the JINGLE public data release to ease comparison with other samples.

\subsection{Derived products catalog}

In addition to the stellar masses and star formation rates described in Sec. \ref{data}, we have compiled and calculated an extensive set of measurements for the JINGLE galaxies, as the survey science objectives revolve around understanding the interplay between gas, dust, and a broad range of galaxy properties. As part of the JINGLE MDR, we release the derived products catalog for all 193 JINGLE galaxie. In addition to JINGLE catalog IDs and SDSS name, coordinates and spectroscopic redshift, the key quantities presented in Table \ref{sampleparams} are: 
\begin{itemize}
\item \mstar: the stellar masses estimated with MAGPHYS and our CAAPR photometric catalog. The median statistical uncertainty on \mstar\ is 0.055 dex and the systematic uncertainty is ${\sim}0.15$ dex, as estimated from the scatter between the MAGPHYS results and other stellar mass estimations as shown in Fig.~\ref{Mstarcomp}. 
\item $r_{50}$: the SDSS $r-$band Petrosian radius, in units of kiloparsec
\item $\mu_{\ast}$: the stellar mass surface density calculated as $\mu_{\ast}=M_{\ast}/(2\pi r_{z}^2)$, where $r_z$ is the Petrosian half-light radius in the $z-$band in units of kiloparsec. This quantity correlates with morphology, with $\log \mu_{\ast}=8.7$ the empirical threshold where galaxies go from being disc- to bulge-dominated.  
\item C: the concentration index defined as the ratio of the SDSS $r-$band Petrosian $r_{90}$ and $r_{50}$. It is a measure of how centrally concentrated the light of the galaxy is with values above 2.5 indicative of a significant stellar bulge contribution to the total light. 
\item M: galaxy morphology as determined from Galaxy Zoo 1 \citep[GZ1;][]{lintott11}, or from KIAS value-added galaxy catalog \citep{choi10} and our own visual classification if not available in GZ1 (1: spiral, 2: elliptical). The vast majority of the galaxies in the JINGLE sample are spirals. Alternative morphology information based on automated classifications or bulge/disc profile fitting, and for example differentiating between early- and late-type spirals, are also available elsewhere \citep[e.g.,][]{huertas11,simard11}.  
\item SFR: the star formation rate obtained with MAGPHYS and the CAAPR photometric catalog.  The median statistical uncertainty on SFR is 0.03 dex and the systematic uncertainty is $\sim0.2$ dex, as estimated from the scatter between the MAGPHYS results and other SFR estimations as shown in Fig.~\ref{SFRcomp}. 
\item 12$+\log$(O/H): gas-phase metallicity calculated from optical strong emission lines measured in the SDSS spectra using the O3N2 calibration of \citet[][hereafter PP04]{pettini04}. In cases where the emission lines are not all detected or where their excitation is likely to be influenced by the presence of an AGN (see column ``BPT''), then we use the value derived from the mass-metallicity relation as derived by \citet{kewley08} to be on the same PP04 scale. 
\item BPT: galaxy classification based on SDSS optical emission line flux ratios using the criteria of \citet{baldwin81}, \citet{kewley01}, and \citet{kauffmann03} (-1: undetermined, 0: inactive, 1: star forming, 2: composite, 3: LINER, 4: Seyfert). The galaxies are not selected in any way based on the presence or not of an active nucleus, and therefore the sample does not contain any bright (and thus rare) AGN, although 14 of the galaxies are classified as LINER or Seyfert.  
\item Env: environment classification based on the information in the group catalog of \citet{tempel14} (0: no data, 1: isolated, 2: central, 3: satellite). 
\end{itemize}

The full version of Table \ref{sampleparams} including all 193 galaxies is available in electronic format and on the JINGLE data release page\footnote{\url{http://www.star.ucl.ac.uk/JINGLE/data.html}}.

\section{JINGLE Main Data Release}
\label{DR}

{\edit Observations for JINGLE at the JCMT began in December 2015, with the SCUBA-2 component of the survey completed in February 2018. Due to particularly good weather conditions throughout the winter of 2016 owing to an El Ni\~{n}o effect, the completion rate of the RxA3m observations, which are designed to be conducted in poorer weather conditions, remained lower. By the time the RxA3m receiver was decommission in June 2018, we had completed observations of \nrxa/\numberCO\ of the intended targets. This completed sample includes all the higher priority MaNGA objects. We therefore include in the JINGLE Main Data Release (MDR) all 193 SCUBA-2 observations and CO(2-1) observations for \nrxa\ of these galaxies. The remaining  galaxies selected for CO observations will be observed as soon as a replacement receiver is installed on the JCMT (expected in 2019) and those data made public in due course in an Extended Data Release. }

\subsection{SCUBA-2}

The SCUBA-2 data are reduced within the Starlink environment \citep{currie14} using a custom-made pipeline for the specificities of the JINGLE observations. Extensive simulations were performed to develop this pipeline, in particular to fully characterize the impact of filtering, and investigations made to find the most appropriate standard flux calibration factor \citep{dempsey13}. Total 850$\mu$m fluxes are measured through aperture photometry, with apertures determined through a joint analysis of the \textit{Herschel-}SPIRE photometry based on the method describe in \citet{Smith2017}. The full details of the SCUBA-2 observations and data reduction are given in \paperII. 

\begin{figure}
    \centering
    \includegraphics[width=3.4in]{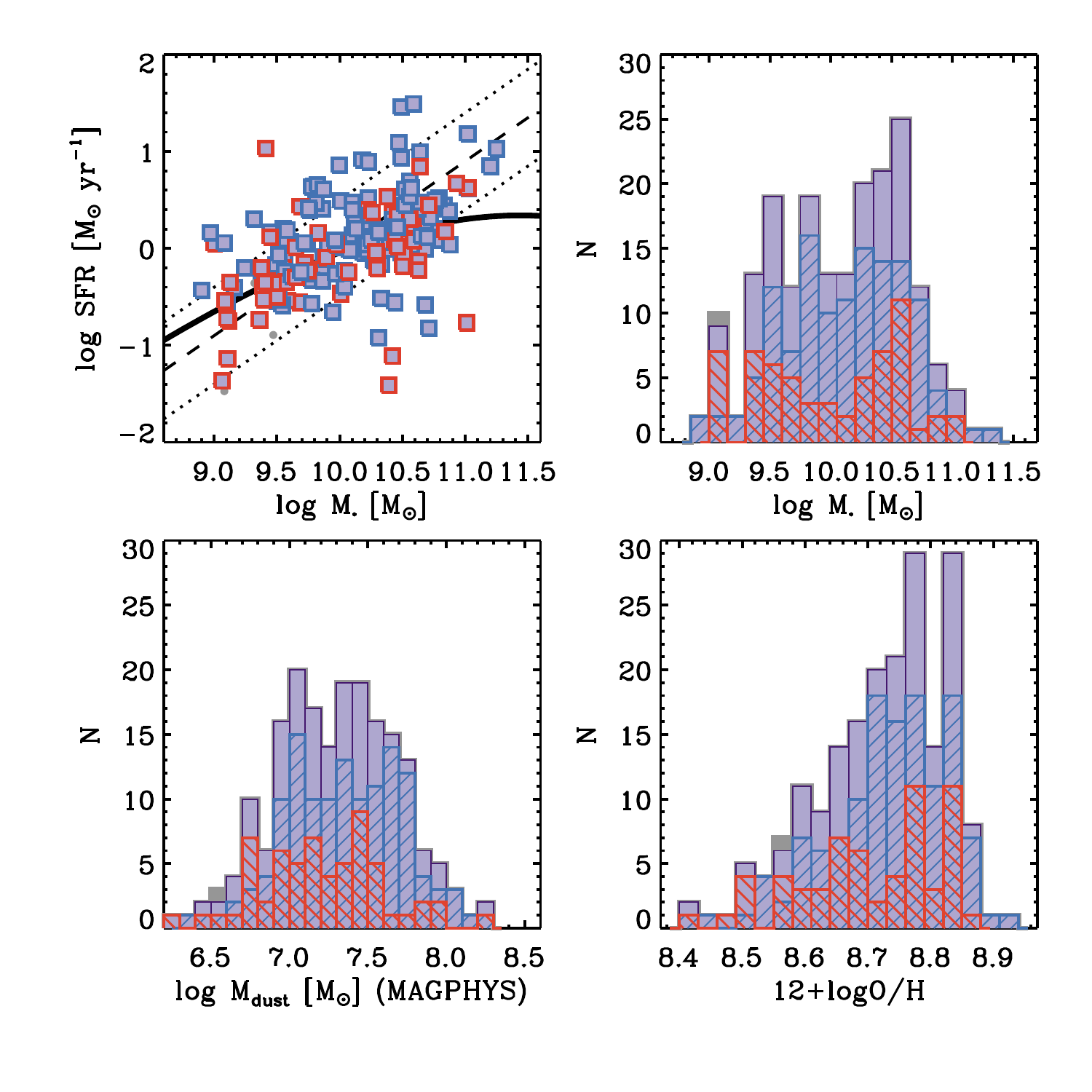}
    \caption{Overview of the SCUBA-2 sample. {\it Top left:} distribution of the JINGLE sample in the SFR-\mstar\ plane.  The points outlined in blue represent the galaxies with a $>3\sigma$ detection of the 850$\mu$m continuum, and red outlines the non-detections. The different lines show the position of the star formation main sequence as in Figures \ref{sampleSCUBA2} and \ref{sampleRxA}. {\it Other three panels:} Histograms showing the distribution of stellar masses, predicted dust masses and metallicities for the full JINGLE sample (filled purple). All these galaxies are included in the MDR. The sample is further shown divided by 850$\mu$m detections (blue) and non-detections (red).}
    \label{SCUBA2status}
\end{figure}

The properties of the sample of galaxies with SCUBA-2 observations is summarised in Figure \ref{SCUBA2status}.  The overall detection rate at 850\micron\ is 64\% (3$\sigma$ detections), but the non-detections do not cluster in any particular region of parameter space.  As part of our MDR, we release the 850$\mu$m maps all 193 JINGLE galaxies with and without matched filtering applied. An example of the 850$\mu$m image of galaxy JINGLE25 is shown in Figure~\ref{SEDexample}. In addition, the MDR catalog presented in \paperII\ includes the fluxes measured from consistent aperture photometry on both our new SCUBA-2 images and the {\it Herschel} PACS and SPIRE images. As explained in Section~\ref{results}, these far-infrared and sub-millimetre measurements are combined to carefully constrain the dust properties of the JINGLE galaxies. 

\subsection{RxA3m}
\begin{figure}
    \centering
    \includegraphics[width=3.4in]{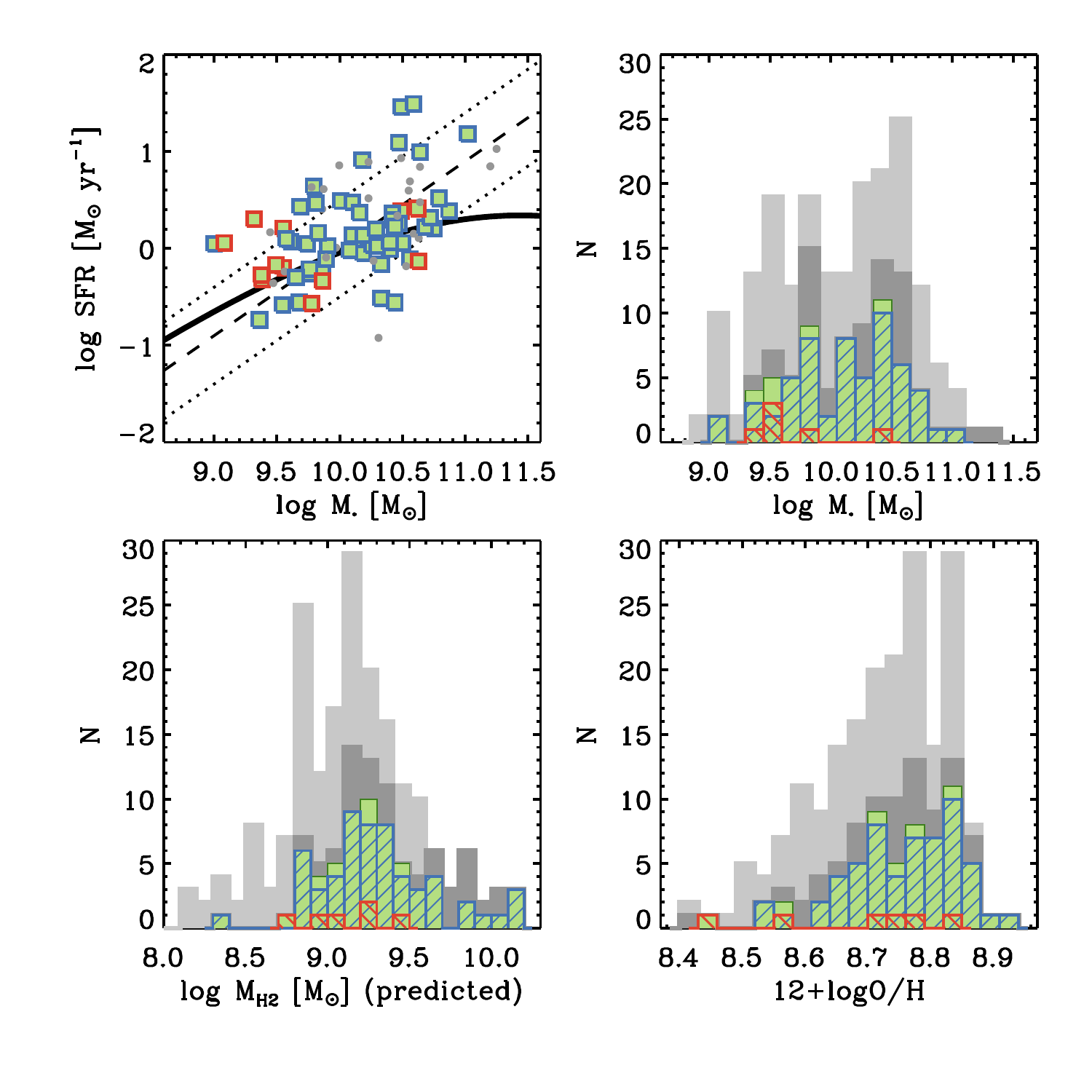}
    \caption{Overview of the RxA3m sample. {\it Top left:} distribution in the \ms\ plane of the sample of \numberCO\ targets for CO(2-1) observations; larger filled squares identify the \nrxa\ galaxies with CO measurements released as part of the JINGLE MDR.  The open blue squares outline the galaxies with a detections of the CO(2-1) line (S/N$>4.5$) and the open red squares the non-detections. {\it Other three panels:} Distribution of stellar masses, predicted molecular gas masses from the 2-SFM formalism \citep{sargent14}, and gas-phase metallicities for the entire JINGLE sample (light gray), the subset of \numberCO\ galaxies to be observed with RxA3m by JINGLE (darker gray), and the CO sample included in the MDR (filled green). The MDR sample is further divided into secure  detections (blue) and more tentative detections (red).}
    \label{fig:rxastatus}
\end{figure}

The status of the RxA3m observations released as part of the MDR in \paperIII\ is summarized in Figure~\ref{fig:rxastatus}.  There are \nrxa\ galaxies with CO observations in MDR. The JINGLE CO sub-sample (dark gray histograms in Fig.~\ref{fig:rxastatus}) is representative of the overall JINGLE sample in terms of stellar mass and metallicity, but biased towards slightly more gas-rich objects, as shown by the distribution of predicted $H_2$ masses.  This selection effect occurs because we include in the CO sub-sample only those galaxies from the full SCUBA-2 sample with a total estimated integration time that is less than 14 hours to reach a 5$\sigma$ detection of the CO(2-1) line. 

In \paperIII, we highlight how the predicted CO(2-1) line luminosities were very accurate, which translates into a high detection rate of 80\%. An example JCMT spectrum for one of the secure detections of the CO(2-1) line (S/N$=8.9$) is shown in Figure~\ref{SEDexample}. The MDR catalog includes the integrated line fluxes and luminosities, molecular gas masses, CO-based redshifts and linewidths for all \nrxa\ galaxies. The linewidths will be used in further studies to improve the calibration of the CO Tully-Fisher relation \citep[e.g.][]{tiley16}. 

\section{Example science}
\label{results}

We present some short highlights of science enabled by JINGLE, all of which will be revisited in more depth in the data release papers and subsequent science analysis papers. 

\subsection{The relation between CO line luminosity and the FIR continuum}

Although measurements of the cold interstellar medium are typically obtained via molecular and atomic line spectroscopy, several recent studies have derived total gas masses via a gas-to-dust ratio combined with far-infrared/sub-mm continuum measurements of total dust masses \citep[e.g.][]{israel97,leroy11,magdis11,eales2012,sandstrom13}. There are also suggestions that the luminosity in particular FIR bands, such as 500 $\mu$m or 850 $\mu$m, could be extrapolated directly to a total molecular gas mass without the need to first estimate a dust mass \citep{scoville14,groves15,scoville16}. These methods are generating significant interest, as they allow gas masses to be measured quickly for very large samples, for example in high-redshift galaxy surveys. Uncertainties related to these methods involve the dependence of the gas-to-dust ratio on metallicity and changes in the physical properties of the dust grains with environment and/or redshift. Dust masses are typically estimated using Milky Way-like dust properties \citep{draineli07} and a simple linear relation between gas-to-dust ratio and metallicity \citep{leroy11}.

\begin{figure}
    \centering
    \includegraphics[width=3.4in]{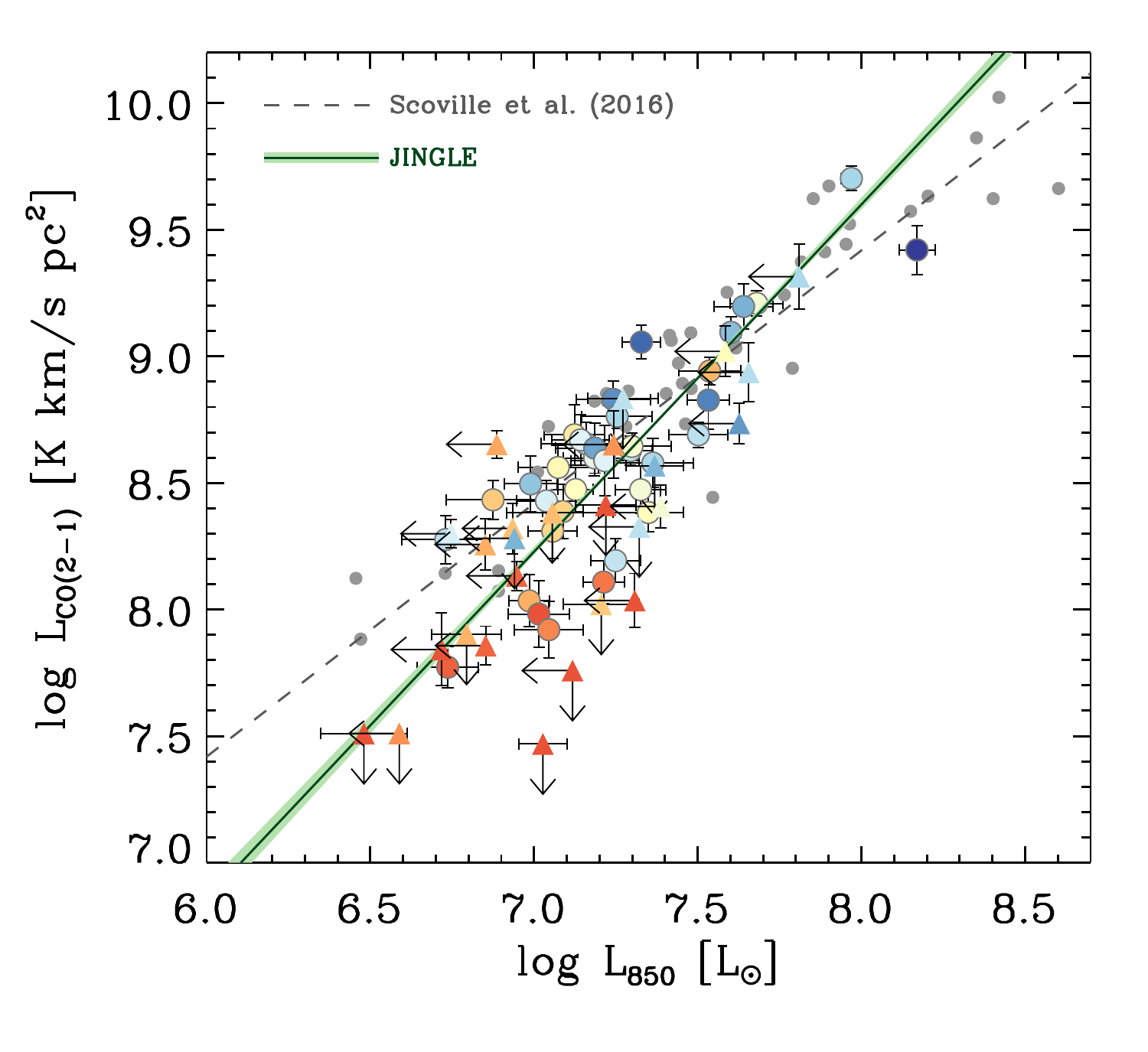}
    \caption{Comparison between the 850\micron\ and CO(2-1) line luminosities of the \nrxa\ JINGLE galaxies with both SCUBA-2 and RxA3m observations in the MDR. Galaxies are colour-coded by stellar mass (red being low mass and dark blue the highest masses). The green solid line and associated shaded error region is the bisector fit to all JINGLE objects, taking into account uncertainties on both axes and all upper limits. For comparison, the reference sample of \citet{scoville16} is shown in gray (after applying a correction of $r_{21}=0.8$), with the best fit relation to this sample shown as the dashed gray line.}
    \label{SRcomp}
\end{figure}

JINGLE will be able to investigate these assumptions and calibrate the empirical relation to estimate gas masses based on FIR/submm continuum. We begin here by investigating the relation between CO(2-1) line luminosity and 850$\mu$m luminosity for those \nrxa\ galaxies in MDR which have both SCUBA-2 and RxA3m observations. Figure~\ref{SRcomp} shows this relation through measuring the 850$\mu$m flux that is coming from the area equivalent to the RxA3m beam at the frequency of the CO(2-1) line. Not surprisingly, there is a clear  and near-linear correlation between the two sets of luminosities, in agreement with the sample compiled by \citet{scoville16}, where we have assumed a CO(2-1)/(1-0) line ratio of $r_{21}=0.8$ \citep{saintonge17} to compare the samples directly.

The relation between 850\micron\ and CO line luminosity calibrated by \citet{scoville16} using a sample of bright nearby star-forming and starburst galaxies is linear in logarithmic space. The JINGLE galaxies as shown in Fig.~\ref{SRcomp} suggest a change in the relationship at the low luminosity end, which is also where the lowest mass (and therefore lowest metallicity) galaxies reside. Fitting to all the galaxies in the JINGLE DR1 sample while carefully accounting for upper limits and measurement errors, we find the relation to be superlinear with $\log L_{CO(2-1)}=1.372\log L_{850}$-1.376. In particular, Fig. \ref{SRcomp} suggests that low mass (and lower metallicity) galaxies are underluminous in CO(2-1) relative to their 850\micron\ emission. Any deviation from a linear dependence or any second parameter dependence in the $L_{CO}-L_{FIR}$ relation will be investigated by JINGLE, and further discussion of the correlations between CO luminosity and monochromatic submillimetre fluxes will be presented in \paperIII.

\subsection{Dust SED modeling}

The new SCUBA-2 850$\mu$m observations, in combination with the ancillary \textit{WISE} 12, 22$\mu$m, \textit{IRAS} 60$\mu$m and \textit{Herschel} 100, 160, 250, 350 and 500$\mu$m data for JINGLE galaxies, result in an exceptionally well-sampled dust spectral energy distribution, extending from the stochastically heated grains probed at mid-infrared wavelengths to the warm and cold dust components emitting in far-infrared and sub-millimetre wavebands. This broad wavelength coverage makes the JINGLE sample a unique laboratory to study the multi-temperature dust reservoirs hosted by galaxies and to probe variations in a galaxy's dust grain properties. To exploit this unique wavelength coverage, we use a set of different types of dust SED models to uncover the nature of grain populations and investigate possible grain property variations with the metallicity, stellar mass, and (specific) star formation rate of JINGLE galaxies. 

\begin{figure*}
\centering
\subfigure{\includegraphics[width=0.45\textwidth]
{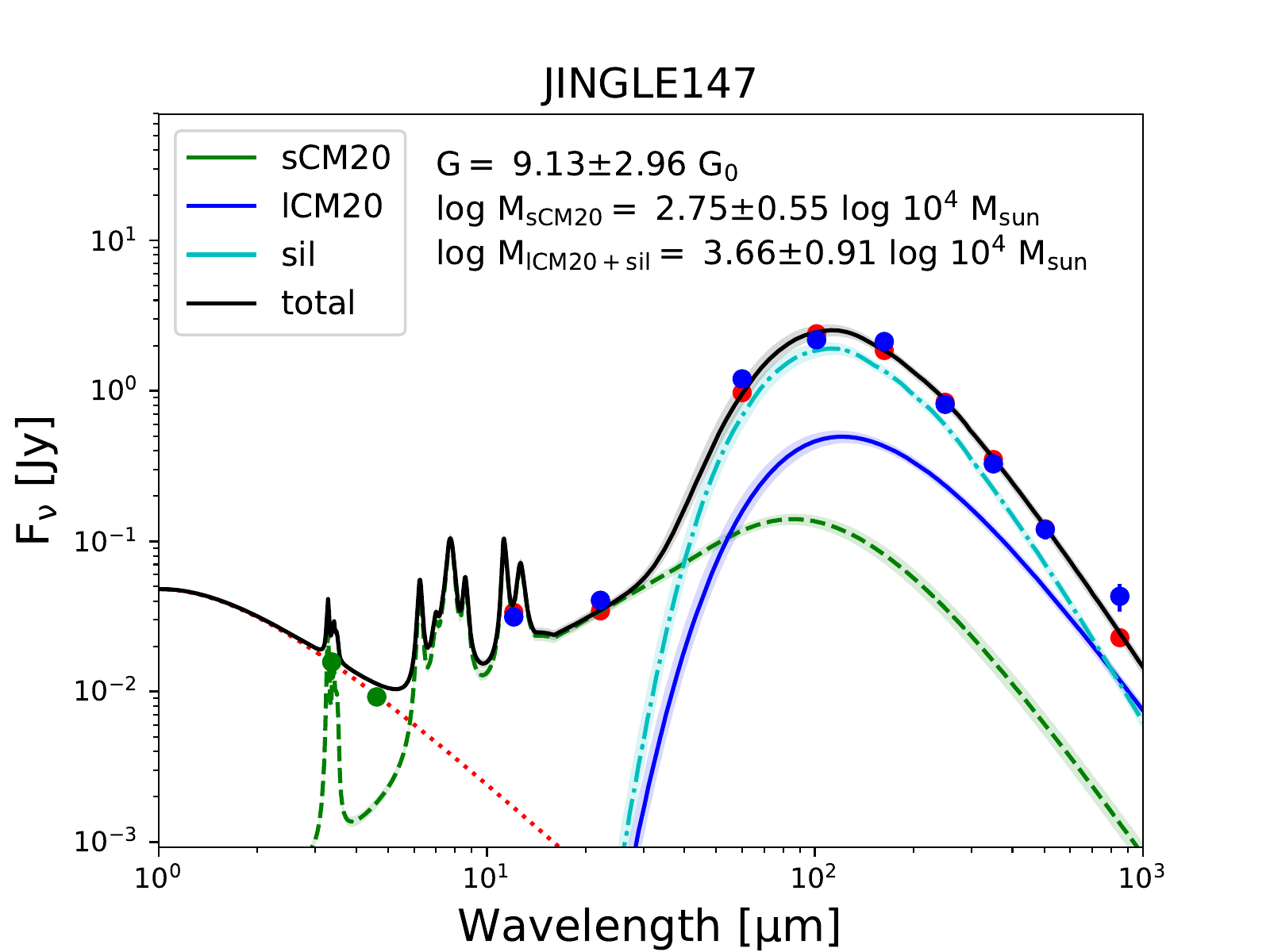}}
\subfigure{\includegraphics[width=0.50\textwidth]
{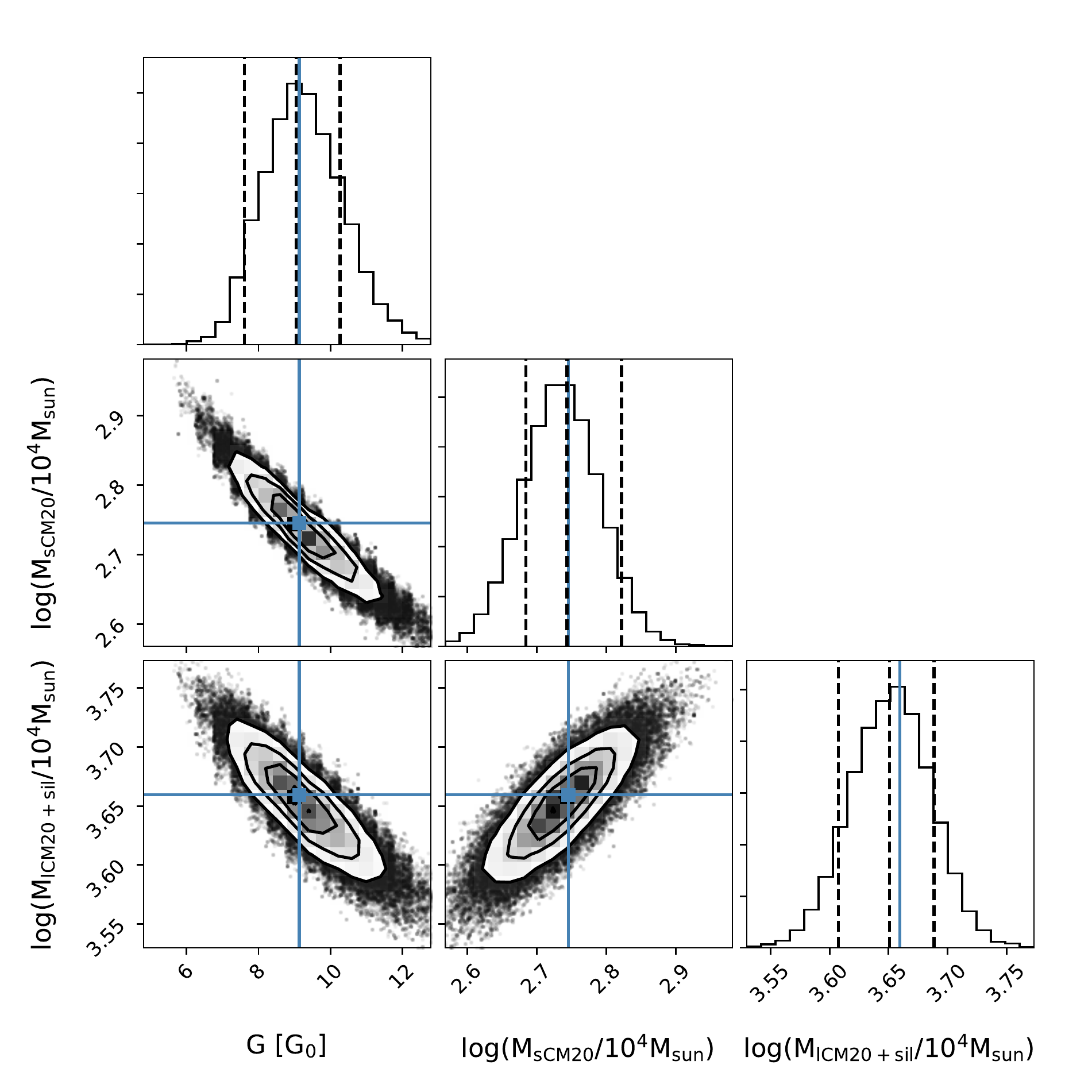}}
\caption{Left panel: example SED for JINGLE\,147, fitted using the THEMIS dust model. The best-fit SED models for small (sCM20) and large (lCM20) carbonaceous grains and large silicate (sil) grains are indicated with green dashed, blue solid and cyan dash-dotted lines, respectively. The stellar emission at NIR wavelengths is modelled using a blackbody function with temperature $T_{\text{d}}$=5,000\,K (red dotted curve). The total best-fit stellar+dust SED emission is shown in black. The shaded regions indicate the lower and upper limit uncertainties on the SED models, as derived from the 16th and 84th percentiles in the posterior distributions. Right panel: probability distribution functions (PDF) which indicate the likelihood of a given output parameter value. The blue line indicates the position of the maximum likelihood (or best-fit model) solution which does not always correspond to the peak of the PDF. 
}
\label{fig:THEMIS_SED_model} 
\end{figure*}

In a first paper (\paperiv), we model the dust emitted from NIR to submm wavebands with The Heterogeneous dust Evolution Model for Interstellar Solids (THEMIS) dust model \citep{Jones2013, Kohler2014, Jones2017}. The THEMIS dust composition consists of hydrogenated amorphous carbons, (a-C(:H)) and silicates with iron nano-particle inclusions (a-Sil$_{\text{Fe}})$. The optical constants for these grain species were derived from laboratory studies and the size distribution and grain abundances were constrained from the observed dust extinction and emission in the Milky Way. We study variations in the relative grain abundances of small (sCM20) and large hydrocarbons (lCM20) and silicate-type grains (sil) across the sample of JINGLE galaxies and determine the strength of the radiation field heating these grains, $G$, relative to the radiation field characteristic of the solar neighbourhood, $G_{\textit{0}}$. Figure \ref{fig:THEMIS_SED_model} shows an example of a best fit SED with the THEMIS dust model for JINGLE\,147, and is representative of the type of modelling applied to the entire JINGLE sample in Paper IV. We will study how the total dust mass and relative grain abundances change depending on whether the SCUBA-2\,850$\mu$m observations are used to constrain the dust SED. We will furthermore present dust scaling relations for the entire JINGLE galaxy sample and compare them with other nearby galaxy samples to infer how ``dusty" JINGLE galaxies are (see also Fig.~\ref{FUVKL12}).

\begin{figure*}
\centering
\subfigure{\includegraphics[width=0.3\textwidth]
{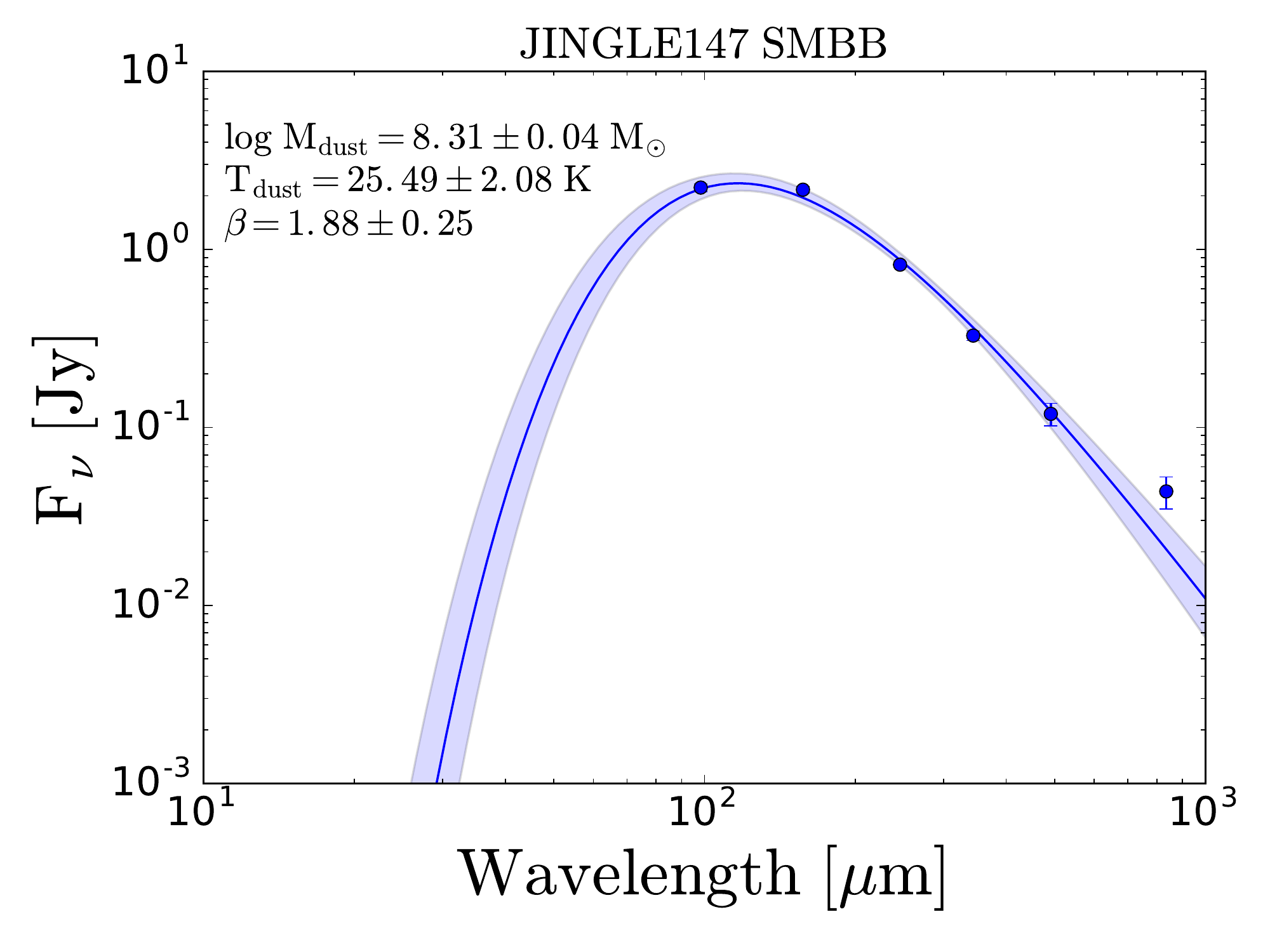}}
\subfigure{\includegraphics[width=0.3\textwidth]
{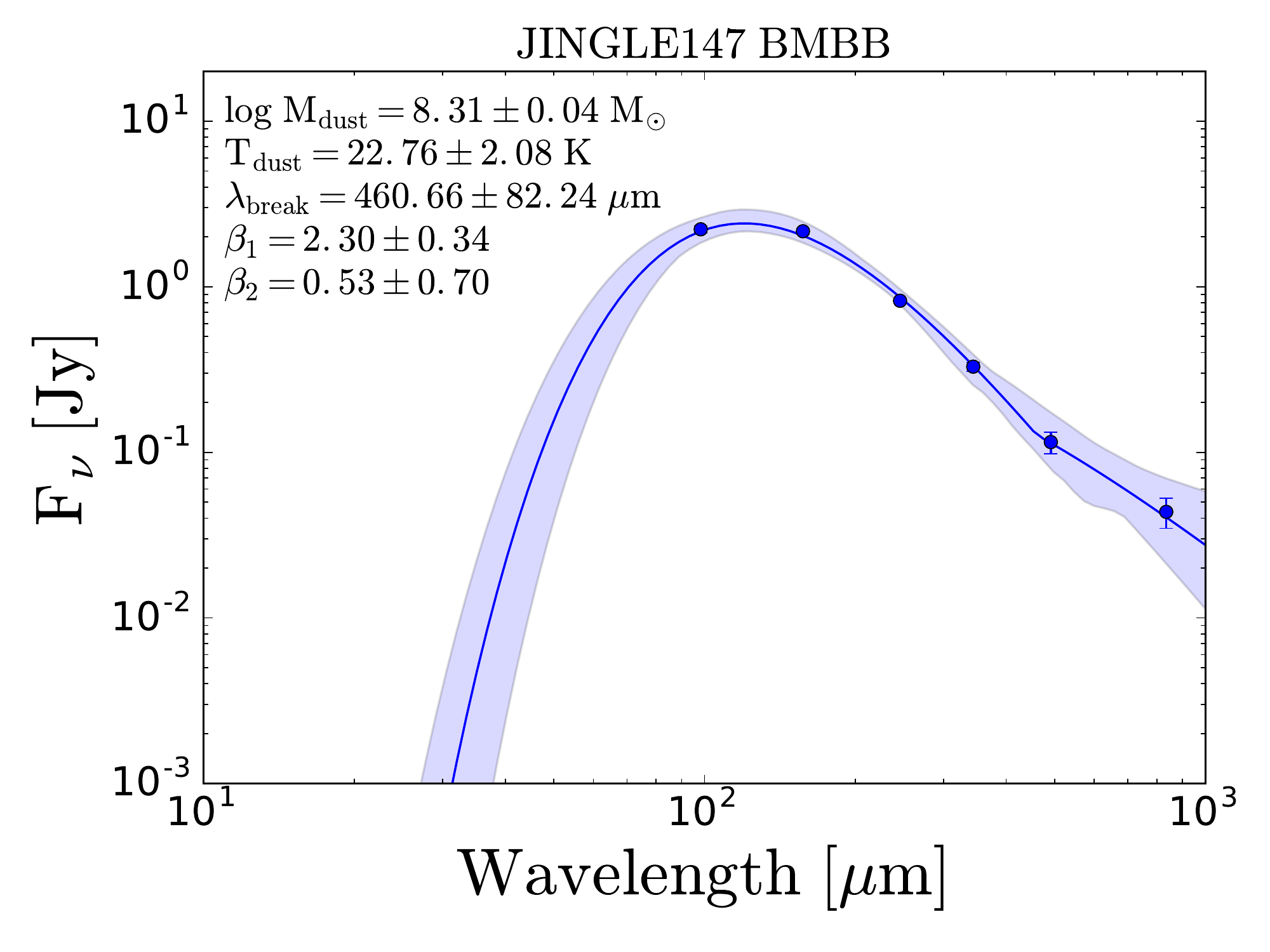}}
\subfigure{\includegraphics[width=0.3\textwidth]
{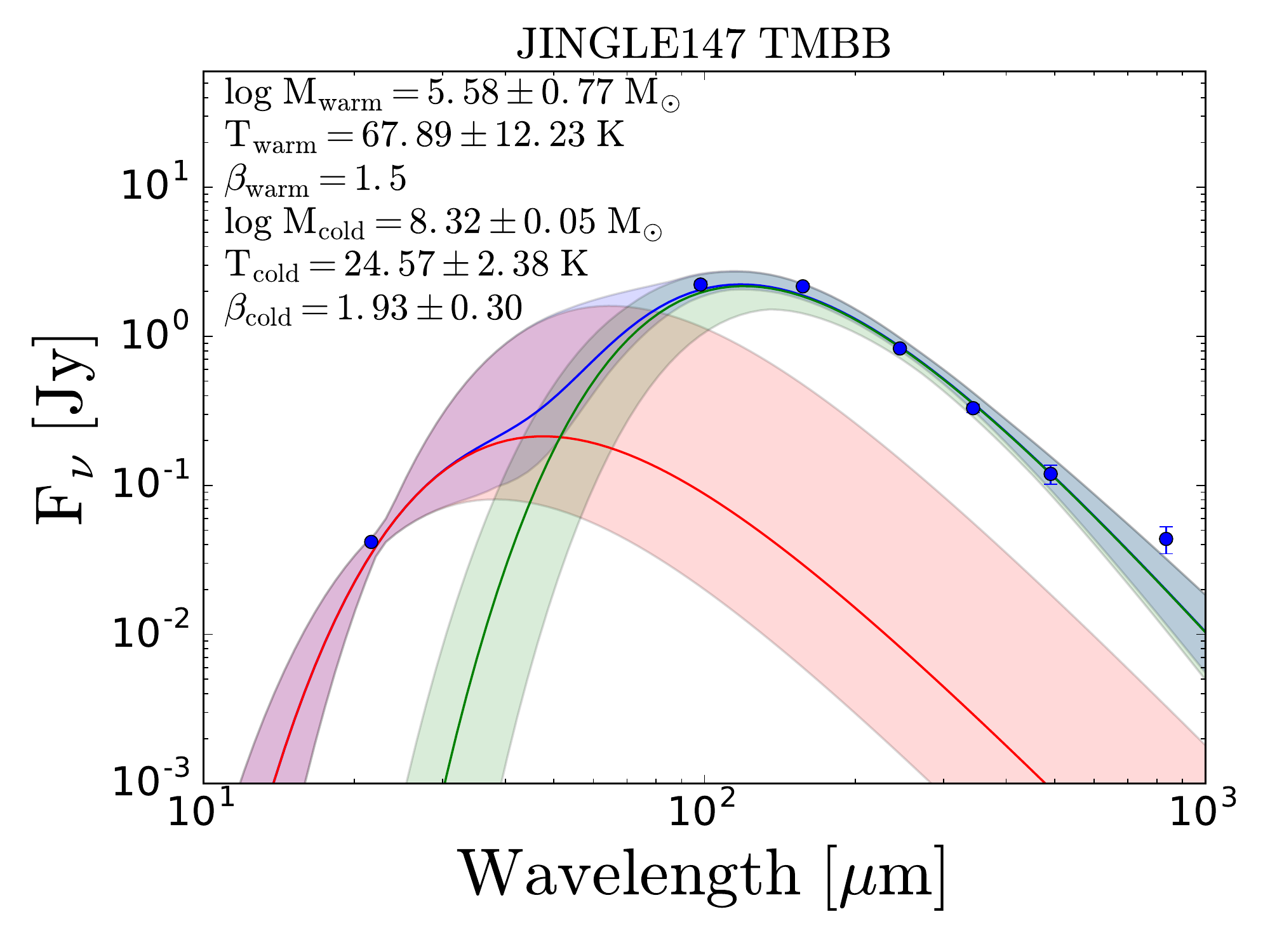}}
\caption{Example SED for JINGLE\,147, fitted using the three models: SMBB (left panel), BMBB (middle panel) and TMBB (right panel). See text for a description of the three models and their parameters.
}
\label{fig:three_SED_models} 
\end{figure*}

In a second paper (Lamperti et al.\,in prep., hereafter Paper V), we model the JINGLE dust emission using a variety of modified blackbody functions (MBBs) to infer how the dust mass, $M_{\text{d}}$, effective dust emissivity index, $\beta_{\text{eff}}$, and dust temperature, $T_{\text{d}}$, vary among the JINGLE sample. The effective dust emissivity index $\beta_{\text{eff}}$ is sensitive to the Rayleigh-Jeans slope of the dust SED and its peak position. The slope depends on the dust emissivity of grains which is directly linked to the composition and size of grains. A Bayesian fitting algorithm is used to derive the best fitting model parameters for a set of different dust SED models. We adopt the three models employed by \citet{Gordon2014} for the SED fit of the Magellanic Clouds: single modified black-body (SMBB), two modified black-bodies (TMBB) and and broken emissivity law modified black-body (BMBB). Figure \ref{fig:three_SED_models} shows representative SED fits using the SMBB, BMBB and TMBB models for JINGLE\,147. We assumed a constant value of $\kappa_0$= $\kappa$(500 $\mu$m$) = 0.051 \text{m kg}^{-1}$ from \citet{Clark2016} in the SED fitting. More details about the dust SED modelling can be found in Paper V. We will also compare non-hierarchical and hierarchical Bayesian fitting algorithms, and study the effect of these different methods on the $T_{\text{d}}$-$\beta$ relation for JINGLE galaxies. The factor of 4 offset in the dust mass derived with the THEMIS dust model and the MBB models for JINGLE\,147 is largely attributed to the different dust opacities assumed in both models, and will be further explored in Papers IV and V. 

\subsection{Background sources}

While the JINGLE SCUBA-2 observations are designed to measure the emission from targeted galaxies, their field of view is significantly larger, allowing for a blind survey of background objects. Over the 193 fields observed as part of JINGLE, the total area mapped by SCUBA-2 is around 10.1 deg$^2$. However, this includes the edges of the maps, which typically have much higher noise than the center, so our fields are not uniform.
We can restrict ourselves to ``good'' pixels by selecting only pixels with instrumental noise resulting in a mean uncertainty of 1.6 mJy beam$^{-1}$ or less, comparable to that seen in the S2 Cosmology Legacy Survey \citep[S2-CLS,][]{geach17}, which covered 2.2 deg$^2$. Under this restriction, the total area covered by JINGLE is 1.05 deg$^2$. The highlight results presented below were however derived from the first 105 fields observed by JINGLE, corresponding to a high sensitivity area of 0.57 deg$^2$. 

To measure the 850\micron\ fluxes, F$_{850}$, of sources other than the main JINGLE targets, we first convolved the maps with a matched filter of 13$^{\prime\prime}$ diameter, equal to the SCUBA-2 beam at 850 $\mu$m.
We then selected all sources with a peak signal to noise ratio of 4 or more in this convolved map and extracted the 850 $\mu$m flux at these positions using aperture photometry on the raw maps.
An aperture of 13$^{\prime\prime}$ radius was used to extract the source flux, with an annulus of inner radius 13$^{\prime\prime}$  and outer radius of 26$^{\prime\prime}$ used to extract a background estimate, which was removed from the source flux.
No further corrections have been made at this stage. 
The positions of the 850 $\mu$m sources were then used to extract sources on the \textit{Herschel} 250, 350 and 500 $\mu$m maps from \textit{H}-ATLAS.
This process results in a total of 119 sources detected across the 105 maps.

\begin{figure}
    \centering
    \includegraphics[width=0.5\textwidth]{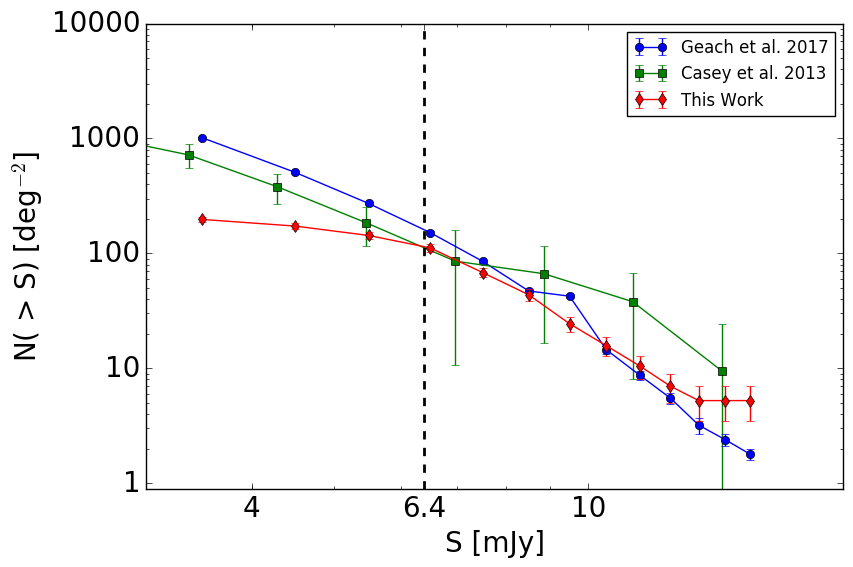}
    \caption{The number counts of sources detected in the JINGLE fields (red diamonds) compared to \citet{geach17} (blue circles) and \citet{casey13} (green squares). The black dashed line gives the 4$\sigma$ detection limit imposed on our sample.}
    \label{fig:NumCounts}
\end{figure}

As a first look, in Figure \ref{fig:NumCounts} we compare our results from all 119 sources to the number counts of 850$\mu$m sources from the $\sim$ 2.2 deg$^2$ S2-CLS and to $\sim$ 0.5 deg$^2$ deep images of the COSMOS field \citep{casey13}.
Even without any correction, we find there is generally good agreement between our observations and the other fields. 
At the high flux end we appear to detect more objects. 
This is to be expected, as our observations target local galaxies as opposed to random fields.
The black dashed line indicates our approximate detection threshold cutoff of 6.4~mJy, and below this we detect fewer sources relative to the blank field number counts, as expected.

\subsubsection{Overdensities of sources}

\begin{figure}
    \centering
    \includegraphics[width=0.5\textwidth]{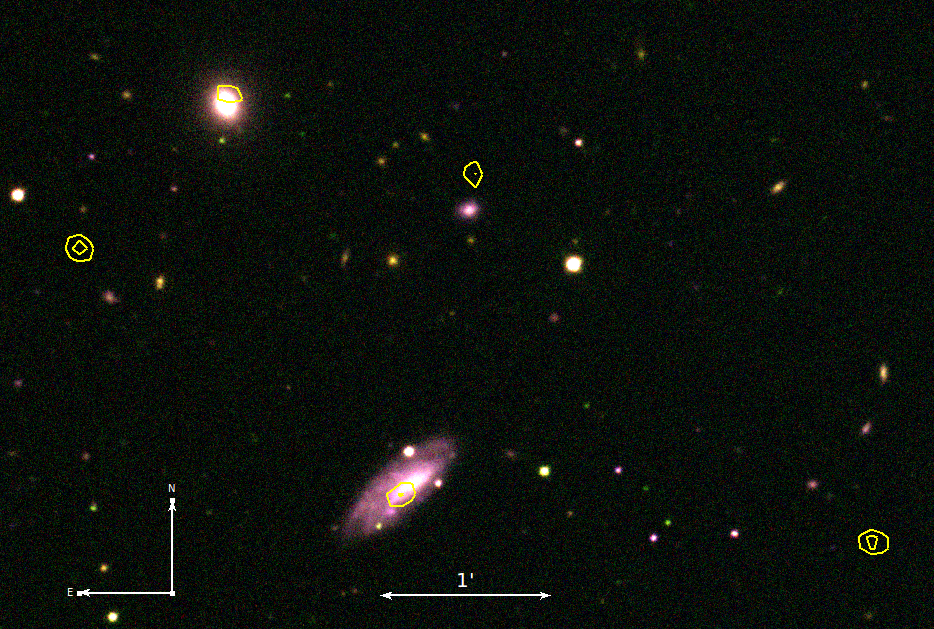}
    \caption{RGB (SDSS $r-$band, SDSS $g-$band, SDSS $i-$band) image of a portion of the JINGLE6 field. The yellow contours show the 4 and 5$\sigma$ detections from SCUBA-2. Three of the detected sources lie outside this field of view. The JINGLE main target galaxy is clearly detected in the bottom-center of the image.}
    \label{fig:Jingle6}
\end{figure}

To focus purely on background sources, we selected all sources that are at a distance of at least 40$^{\prime\prime}$ (approximately three times the FWHM of the SCUBA-2 beam) from the central galaxy that was targeted.
Of our 119 sources, 79 fulfil  this criteria.  In the JINGLE6 field, we detect 8 SCUBA-2 sources to at least a 4$\sigma$ level, some of which are shown in Figure \ref{fig:Jingle6} . 
Their fluxes vary between 3.7 and 7.5 mJy, with a mean of $6.0 \pm 1.3$ mJy.
One is associated with the central galaxy and one appears to be associated with the $z = 0.0159$ galaxy 2MASX J13232557+3206115,  but the other 6 do not appear to be associated with any optical source.

Using the 850$\mu$m number counts from \citet{geach17}, and counting those sources with a 850$\mu$m flux greater than 6.3 mJy, we expect to detect $\sim 175.8 \pm 4.7$ sources per square degree.
In the 0.02 deg$^2$ of JINGLE6, we detect 5 sources with an 850$\mu$m flux greater than 6.3 mJy, two of which are associated with local galaxies.
Converting this to a number counts estimate (without corrections) would result in 252.1 $\pm$ 15.9 sources deg$^{-2}$, a 4.8$\sigma$ over-density. 
Given that two of our sources appear to be associated with local galaxies, this is unlikely to be a physical cluster of 850 $\mu$m sources, and is more likely to be merely a line of sight over-density. 
We note, however, that 2MASX J13232557+3206115 is classed as an elliptical galaxy, and is unlikely to have a significant infrared flux. 
No source is detected at this position in the \textit{Herschel} 250, 350 or 500$\mu$m maps, though a $\sim$ 2$\sigma$ 500$\mu$m flux of 12~mJy does appear $\sim$ 10\arcsec\ away from the nominal position, within the size of SPIRE's 500$\mu$m beam.
It is therefore possible that this source is lensing a background source or that there is a chance overlap between this local galaxy and a background SMG.
Other fields, such as JINGLE21 and JINGLE91, also show mild over-densities at a few $\sigma$ level, but JINGLE6 appears to be the most over-dense of the 105 JINGLE fields studied for background sources so far. 

\subsubsection{Quasars}

The quasar B2 1310+31 at $z=1.055$ \citep{colla70} is detected to a 21$\sigma$ level with F$_{850}$ = 24.6$\pm$1.2 mJy.
This source is not detected in the \textit{Herschel} 250, 350 or 500$\mu$m maps.
A non-detection with \textit{Herschel} is not surprising, as the source has a reasonably flat spectrum and \textit{Herschel}'s detection limit is around 20 mJy. The 850$\mu$m flux is consistent with the radio flux of this flat spectrum source. 

\subsubsection{850$\mu$m risers: high redshift candidates?} 

\citet{michalowski17} find that 20-25\% of  850$\mu$m sources with S/N $\geq$ 4 are not identified in other bands.
In their estimate of the redshifts of these sources, they find they are typically at z $>$ 2, greater than those of 850$\mu$m sources with counterparts in optical/NIR or \textit{Herschel} bands \citep[c.f. Fig.~6 of ][]{michalowski17}.

\begin{figure}
    \centering
    \includegraphics[width=0.5\textwidth]{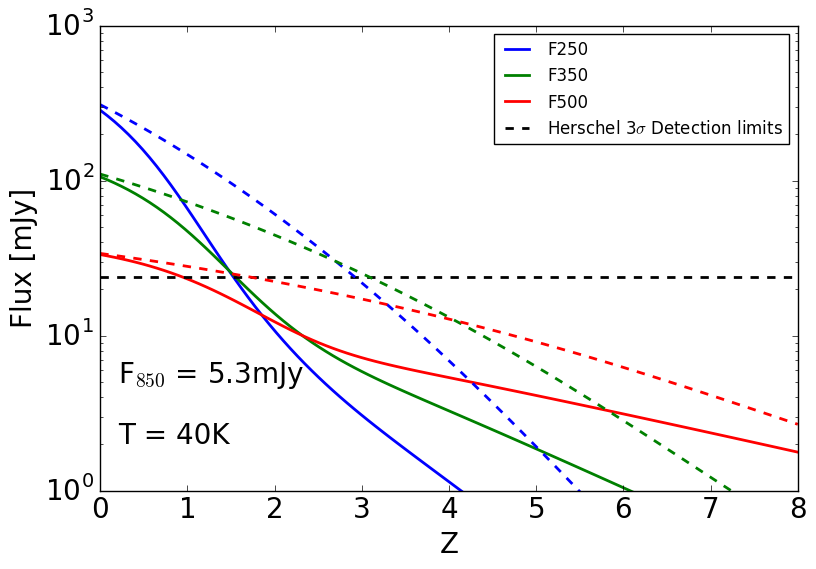}
    \caption{The predicted \textit{Hershel} 250 (blue), 350 (green) and 500 $\mu$m (red) flux of a source with F$_{850}$ = 5.3mJy as a function of redshift. This model assumes a single dust temperature modified blackbody. Solid lines indicate a source that becomes optically thick at 100$\mu$m, and dashed lines indicate a source that becomes optically thick at 10$\mu$m. The horizontal black dashed line indicates the approximate 3$\sigma$ detection limit of \textit{Herschel}.}
    \label{fig:bbsim}
\end{figure}

In our sample of 119 sources, 26 (22\%) have no counterpart in any of the three \textit{Herschel} bands to at least a 3$\sigma$ level.
This is in good agreement with the results of \citet{michalowski17}.
The mean F$_{850}$ for these sources with no counterpart is $5.3 \pm 1.1$ mJy.
To examine the likely redshifts of these sources, we simulate at what redshifts we could reasonably expect to detect a 250, 350 or 500 $\mu$m \textit{Herschel} detection by simulating the FIR flux using the single dust temperature modified blackbody function
\begin{equation}
\label{Eq:BB}
S_\nu \propto \Big(\frac{\nu}{\nu_0}\Big)^{\beta} B_\nu(T),
\end{equation}
typically used to model FIR SEDs (Kelly et al. 2012). Here $\Big(\frac{\nu}{\nu_0}\Big)^{\beta}$ is the opacity of the dust, $\nu_0$ is the characteristic frequency at which the dust becomes optically thick, $\beta$ is the dust emissivity and $B_\nu(T)$ is the Planck function at temperature T.
We assume a dust temperature of 40 K and two assumptions for the dust optical depth: a source with an optical depth that approaches 1 at 10 $\mu$m and a source with an optical depth that approaches 1 at 100 $\mu$m.
We then fixed the 850 $\mu$m flux to the mean flux in our sample of unidentified sources and our results are shown in Figure \ref{fig:bbsim}.
We find that, if these sources are at z < 2, we would reasonably expect to detect them, at least in the 250 $\mu$m band.
As we do not detect any SPIRE flux from these sources, it is difficult to constrain their properties much further, but Fig. \ref{fig:bbsim} indicates that our average SPIRE-dropout is at least consistent with being a population of low luminosity SMGs at z = 2

If we relax our constraint that the mean uncertainty be less than 1.6~mJy, we can search for rarer objects by increasing the area we are examining.
In JINGLE101, we detect a background source to a 4.3$\sigma$ level, with a 850$\mu$m flux of $18.9 \pm 4.9$~mJy. This source is not detected in the 450$\mu$m maps, nor is it detected to a significant level in the \textit{Herschel} 250, 350 or 500 $\mu$m maps. 
At best, it is detected to a 2.6$\sigma$ level in the 500$\mu$m map, with a flux of $12.0 \pm 4.7$~mJy.
Repeating the above blackbody simulation for this source suggests that, if its true 850$\mu$m flux is 18.9 mJy, we would expect to detect it in the 500$\mu$m band out to at least z = 5.5, assuming a dust temperature of 40K and the source being optically thick at 100 $\mu$m.
Assuming an optical depth of 1 at 1$\mu$m, we should expect to detect this source out to z = 7.
The nature of this {\it Herschel} dropout is uncertain: it could be a higher redshift analog of the 500 $\mu$m risers (F$_{250} < $F$_{350} < $F$_{500}$), typically the highest redshift SMGs discovered by \textit{Herschel}, or it could be part of a lower redshift but cooler population of SMGs, with dust temperatures below those of typical dusty star-forming galaxies at these redshifts.
However, the uncertainty on this source is somewhat large, and in fact lies outside of our initial selection limit of 3 mJy. 
We have been allocated ALMA time at 2mm to further constrain the nature of this and several other bright 850 $\mu$m risers in the JINGLE fields, the results of which will be presented in a future paper.

\section{Conclusions}
\label{conclusion}

We have introduced JINGLE, an ongoing large programme at the James Clerk Maxwell Telescope, and its Main Data Release (MDR). The survey is designed to systematically study the cold ISM of galaxies in the local Universe. Over the period of 2015-2019, and making use of 780 hours of observing time on the JCMT, JINGLE will provide integrated 850$\mu$m continuum measurements with SCUBA-2 for a representative sample of \ntot\ {\it Herschel}-selected galaxies, as well as  CO(2-1) line fluxes and spectra with RxA3m for a subset of \numberCO\ of these galaxies. The galaxies in the sample have redshifts $0.01<z<0.05$ and stellar masses in the range $10^9-10^{11.5}$\msun. They are selected in SDSS from four fields chosen for having {\it Herschel} H-ATLAS imaging as well as coverage by the MaNGA and SAMI integral field optical spectroscopy surveys and upcoming large area blind HI synthesis surveys. 

The JCMT observations will allow for the robust characterization of the dust properties (e.g., temperature, emissivity, grain properties) as well as the measurement of total molecular gas masses for the RxA3m subsample. The combination of all these datasets will allow a detailed characterisation of the gas and dust properties and of the kinematics and metal contents of these galaxies, the derivation of scaling relations between dust, gas, and global properties, as well as provide critical benchmarks for high-redshift studies with JCMT and ALMA.

The Main Data Release includes the SCUBA-2 observations for all \nscuba\ JINGLE galaxies, and RxA3m CO(2-1) line measurements for a subset of \nrxa\ of those. In addition, we have produced and release here a 30-band matched-aperture multi-wavelength catalog, including fluxes from {\it GALEX} FUV up to {\it Herschel} 500$\mu$m. This catalog is used to measure accurate and homogeneous stellar masses, star formation rates and total infrared luminosities to be used alongside the JCMT data products. 
 
Based on the \nrxa\ MDR galaxies with observations of both the CO(2-1) line and the 850$\mu$m continuum, we show how low mass galaxies (\mstar$<10^{10}$\msun) steepen the slope of the relation between $L_{CO}$ and $L_{850}$ and increase its scatter. By also quantifying how the properties of dust vary across the galaxy population, one of the aims of the survey is to calibrate how such relations can be used to infer the cold gas mass of galaxies with low metallicities and/or at high redshifts. In the three other papers accompanying this data release, we present in detail the RxA3m and SCUBA-2 observations as well as the catalogs of CO(2-1) line fluxes and sub-millimetre continuum measurements, and present some of the first scaling relations between dust properties and global galaxy properties.

\section*{Acknowledgements}

The James Clerk Maxwell Telescope is operated by the East Asian Observatory on behalf of The National Astronomical Observatory of Japan, Academia Sinica Institute of Astronomy and Astrophysics, the Korea Astronomy and Space Science Institute, the National Astronomical Observatories of China and the Chinese Academy of Sciences (Grant No. XDB09000000), with additional funding support from the Science and Technology Facilities Council of the United Kingdom and participating universities in the United Kingdom and Canada. Additional funds for the construction of SCUBA-2 were provided by the Canada Foundation for Innovation. This data is being observed under JCMT Project ID: M16AL005. The Starlink software \citep{currie14} used as part of the JINGLE data reduction process is currently supported by the East Asian Observatory.

The authors wish to recognize and acknowledge the very significant cultural role and reverence that the summit of Mauna Kea has always had within the indigenous Hawaiian community.  We are most fortunate to have the opportunity to conduct observations from this mountain.

A.S.~acknowledges support from the Royal Society through the award of a University Research Fellowship. 
C.D.W.~acknowledges support from the Natural Sciences and Engineering Research Council of Canada. 
T.X.~acknowledges the support of National Science Foundation of China (NSFC), Grant No. 11203056. 
C.J.R.C.~acknowledges support from European Research Council (ERC) in the form of the 7\textsuperscript{th} Framework Program (FP7) project DustPedia (PI Jon Davies, proposal 606824).
D.L.C.~is supported through STFC grants ST/N000838/1, ST/K001051/1 and ST/N005317/1.
I.D.L~gratefully acknowledges the support of the Science and Technology Facilities Council (STFC) and the Flemish Fund for Scientific Research (FWO-Vlaanderen).
T.A.D.~acknowledges support from a STFC Ernest Rutherford Fellowship. 
M.B.~was supported by the consolidated grants ``Astrophysics at Oxford" ST/N000919/1 and ST/K00106X/1 from STFC. 
L.C.H.~was supported by the National Key R\&D Program of China (2016YFA0400702) and NSFC (11473002, 11721303)
C.L.~acknowledges the support by National Key Basic Research Program of China (2015CB857004) and NSFC (11233005, 11325314, 11320101002).
J.M.S.~acknowledges support from STFC, grant number ST/L000652/1.
M.J.M.~acknowledges the support of  the National Science Centre, Poland through the POLONEZ grant 2015/19/P/ST9/04010; this project has received funding from the EU's Horizon 2020 research and innovation programme under the Marie Sk{\l}odowska-Curie grant agreement No. 665778.
M.T.S.~was supported by a Royal Society Leverhulme Trust Senior Research Fellowship (LT150041).
H.L.G. and P.J.C acknowledge support from the European Research Council (ERC) through the Consolidator Grant {\sc CosmicDust} (ERC-2014-CoG-647939, PI H\,L\,Gomez).
E.B.~acknowledges support from STFC, grant number ST/M001008/1.
J.H.~is supported by the Yunnan Province of China (No.2017HC018).
C.Y., Y.G., X.J. and Q.J. acknowledge support by the National Key R\&D Program of China (2017YFA0402700), the CAS Key Research Program of Frontier Sciences, and the NSFC grants (11311130491, 11420101002).
M.Z.~acknowledges the support by the National Key R\&D Program of China (2017YFA0402600) and by NSFC (U1531246).

This research has made use of Astropy\footnote{\url{http://www.astropy.org/}}, a community-developed core Python package for Astronomy \citep{Astropy2013}. This research has made use of TOPCAT\footnote{\url{http://www.star.bris.ac.uk/~mbt/topcat/}} \citep{Taylor2005A}, which was initially developed under the UK Starlink project, and has since been supported by PPARC, the VOTech project, the AstroGrid project, the AIDA project, the STFC, the GAVO project, the European Space Agency, and the GENIUS project. This research has made use of NumPy\footnote{\url{http://www.numpy.org/}} \citep{VanDerWalt2011B}, SciPy\footnote{\url{http://www.scipy.org/}}, and MatPlotLib\footnote{\url{http://matplotlib.org/}} \citep{Hunter2007A}. This research made use of APLpy\footnote{\url{https://aplpy.github.io/}}, an open-source plotting package for Python \citep{Robitaille2012B}. This research has made use of the scikit-image\footnote{\url{http://scikit-image.org/}} image analysis library \citep{Scikit-Image2014}.

This research made use of {\sc Montage}\footnote{\url{http://montage.ipac.caltech.edu/ }}, which is funded by the National Science Foundation under Grant Number ACI-1440620, and was previously funded by the National Aeronautics and Space Administration's Earth Science Technology Office, Computation Technologies Project, under Cooperative Agreement Number NCC5-626 between NASA and the California Institute of Technology.

This research has made use of GALEX data obtained from the Mikulski Archive for Space Telescopes (MAST); support for MAST for non-HST data is provided by the NASA Office of Space Science via grant NNX09AF08G and by other grants and contracts (MAST is maintained by STScI, which is operated by the Association of Universities for Research in Astronomy, Inc., under NASA contract NAS5-26555). 

This research has made use of data from the 3\textsuperscript{rd} phase of the Sloan Digital Sky Survey (SDSS-III). Funding for SDSS-III has been provided by the Alfred P. Sloan Foundation, the Participating Institutions, the National Science Foundation, and the U.S. Department of Energy Office of Science. The SDSS-III web site is \url{http://www.sdss3.org/}. SDSS-III is managed by the Astrophysical Research Consortium for the Participating Institutions of the SDSS-III Collaboration including the University of Arizona, the Brazilian Participation Group, Brookhaven National Laboratory, Carnegie Mellon University, University of Florida, the French Participation Group, the German Participation Group, Harvard University, the Instituto de Astrofisica de Canarias, the Michigan State/Notre Dame/JINA Participation Group, Johns Hopkins University, Lawrence Berkeley National Laboratory, Max Planck Institute for Astrophysics, Max Planck Institute for Extraterrestrial Physics, New Mexico State University, New York University, Ohio State University, Pennsylvania State University, University of Portsmouth, Princeton University, the Spanish Participation Group, University of Tokyo, University of Utah, Vanderbilt University, University of Virginia, University of Washington, and Yale University. 

This research has made use of the NASA SkyView\footnote{\url{http://skyview.gsfc.nasa.gov/current/cgi/query.pl}} service. SkyView has been developed with generous support from the NASA AISR and ADP programs (P.I. Thomas A. McGlynn) under the auspices of the High Energy Astrophysics Science Archive Research Center (HEASARC) at the NASA/ GSFC Astrophysics Science Division. 

This research makes use of data products from the Two Micron All Sky Survey, which is a joint project of the University of Massachusetts and the Infrared Processing and Analysis Center/California Institute of Technology, funded by the National Aeronautics and Space Administration and the National Science Foundation.

This research makes use of data products from the Wide-field Infrared Survey Explorer, which is a joint project of the University of California, Los Angeles, and the Jet Propulsion Laboratory/California Institute of Technology, and NEOWISE, which is a project of the Jet Propulsion Laboratory/California Institute of Technology. WISE and NEOWISE are funded by the National Aeronautics and Space Administration.

This work is based in part on observations made with the {\it Spitzer} Space Telescope, which is operated by the Jet Propulsion Laboratory, California Institute of Technology under a contract with NASA.

{\it Hershcel} is an ESA space observatory with science instruments provided by European-led Principal Investigator consortia and with important participation from NASA. The {\it Hershcel} spacecraft was designed, built, tested, and launched under a contract to ESA managed by the {\it Hershcel}/{\it Planck} Project team by an industrial consortium under the overall responsibility of the prime contractor Thales Alenia Space (Cannes), and including Astrium (Friedrichshafen) responsible for the payload module and for system testing at spacecraft level, Thales Alenia Space (Turin) responsible for the service module, and Astrium (Toulouse) responsible for the telescope, with in excess of a hundred subcontractors.

\bibliographystyle{mnras}
\bibliography{refs_jcmt}

\begin{landscape}
\begin{table}
\caption{Properties of the JINGLE galaxies (the full table is available electronically) \label{sampleparams}}
\begin{tabular}{lccccccccccccc}
\hline
JINGLE ID & SDSS name & $\alpha_{J2000}$ & $\delta_{J2000}$ &$z_{spec}$ & $\log M_{\ast}$ & $r_{50}$ & $\log \mu_{\ast}$ &  C & M &  
$\log$SFR & 12$+\log$(O/H) & BPT & Env  \\ 
 & & [deg]  & [deg] &  &  $[M_{\odot}]$ & [kpc] & $[M_{\odot}{\rm kpc}^{-2}]$ &  & &   
$[M_{\odot}{\rm yr}^{-1}]$ &  & & \\
\hline
  JINGLE0 &  J131616.82+252418.7 & 199.07012 &  25.40522 & 0.0129 & $10.31\pm0.08$ &  3.78 &  9.15 &  2.78 &  1 & $-0.92\pm0.05$ &  8.75 &  3 &  2 \\
  JINGLE1 &  J131453.43+270029.2 & 198.72264 &  27.00812 & 0.0154 & $ 9.95\pm0.10$ &  5.70 &  8.47 &  2.78 &  1 & $-0.66\pm0.12$ &  8.78 &  1 &  1 \\
  JINGLE2 &  J131526.03+330926.0 & 198.85848 &  33.15724 & 0.0162 & $ 9.12\pm0.12$ &  3.44 &  8.11 &  2.57 &  1 & $-0.75\pm0.06$ &  8.64 &  1 &  1 \\
  JINGLE3 &  J125606.09+274041.1 & 194.02541 &  27.67810 & 0.0165 & $ 9.00\pm0.01$ &  2.23 &  8.10 &  2.44 &  1 & $ 0.05\pm0.02$ &  8.56 &  1 &  3 \\
  JINGLE4 &  J132134.91+261816.8 & 200.39549 &  26.30467 & 0.0165 & $ 9.86\pm0.05$ &  2.73 &  8.95 &  2.63 &  1 & $-0.26\pm0.02$ &  8.82 &  1 &  1 \\
  JINGLE5 &  J091728.99-003714.1 & 139.37082 &  -0.62058 & 0.0166 & $ 9.97\pm0.07$ &  7.09 &  8.37 &  2.59 &  1 & $ 0.01\pm0.02$ &  8.76 &  1 &  3 \\
  JINGLE6 &  J132320.14+320349.0 & 200.83396 &  32.06361 & 0.0167 & $ 9.49\pm0.08$ &  6.00 &  7.85 &  2.25 &  1 & $-0.54\pm0.04$ &  8.68 &  1 &  3 \\
  JINGLE7 &  J132051.75+312159.8 & 200.21563 &  31.36661 & 0.0168 & $ 9.55\pm0.04$ &  5.13 &  8.03 &  2.44 &  1 & $-0.58\pm0.05$ &  8.68 &  1 &  3 \\
  JINGLE8 &  J091642.17+001220.0 & 139.17575 &   0.20556 & 0.0169 & $ 9.68\pm0.07$ &  3.29 &  8.90 &  2.69 &  1 & $-0.56\pm0.05$ &  8.65 &  2 &  1 \\
  JINGLE9 &  J131547.11+315047.1 & 198.94630 &  31.84642 & 0.0170 & $ 9.86\pm0.18$ &  5.87 &  8.07 &  2.36 &  1 & $ 0.41\pm0.23$ &  8.68 &  1 &  2 \\
 JINGLE10 &  J091750.80-001642.5 & 139.46168 &  -0.27848 & 0.0175 & $10.45\pm0.05$ &  7.98 &  8.78 &  2.41 &  1 & $ 0.01\pm0.01$ &  8.72 &  1 &  2 \\
 JINGLE11 &  J131020.14+322859.4 & 197.58392 &  32.48319 & 0.0176 & $ 9.75\pm0.06$ &  9.16 &  7.94 &  2.42 &  1 & $-0.25\pm0.02$ &  8.65 & -1 &  1 \\
 JINGLE12 &  J132251.07+314934.3 & 200.71281 &  31.82622 & 0.0178 & $ 9.38\pm0.05$ &  6.49 &  7.68 &  2.15 &  1 & $-0.32\pm0.02$ &  8.62 &  1 &  1 \\
 JINGLE13 &  J114253.92+000942.7 & 175.72470 &   0.16187 & 0.0185 & $ 8.97\pm0.01$ &  3.02 &  8.11 &  2.25 &  1 & $ 0.16\pm0.29$ &  8.49 &  1 &  3 \\
 JINGLE14 &  J131721.28+310334.1 & 199.33871 &  31.05948 & 0.0186 & $ 9.38\pm0.04$ &  3.93 &  8.09 &  2.38 &  1 & $-0.28\pm0.04$ &  8.56 &  1 &  3 \\
 JINGLE15 &  J090655.44-000152.7 & 136.73103 &  -0.03131 & 0.0187 & $ 9.55\pm0.00$ &  0.00 &  0.00 &  0.00 &  1 & $ 0.21\pm0.00$ &  8.83 &  2 &  2 \\
 JINGLE16 &  J131620.53+304042.0 & 199.08556 &  30.67834 & 0.0189 & $ 9.87\pm0.00$ &  6.09 &  8.55 &  2.35 &  1 & $-0.09\pm0.00$ &  8.62 &  2 &  3 \\
 JINGLE17 &  J130802.57+271840.0 & 197.01072 &  27.31113 & 0.0196 & $ 9.16\pm0.17$ &  3.15 &  8.05 &  2.78 &  1 & $-0.40\pm0.12$ &  8.60 &  1 &  1 \\
 JINGLE18 &  J120803.96+004151.2 & 182.01651 &   0.69758 & 0.0197 & $ 8.90\pm0.00$ &  3.85 &  7.74 &  2.12 &  1 & $-0.43\pm0.00$ &  8.53 &  1 &  1 \\
 JINGLE19 &  J125818.23+290743.6 & 194.57600 &  29.12878 & 0.0263 & $10.63\pm0.01$ &  7.49 &  8.79 &  2.92 &  1 & $-0.13\pm0.05$ &  8.69 &  2 &  3 \\
 JINGLE20 &  J130316.24+280149.4 & 195.81769 &  28.03040 & 0.0203 & $10.39\pm0.03$ &  5.96 &  8.90 &  2.68 &  1 & $-1.41\pm0.15$ &  8.76 & -1 &  3 \\
 JINGLE21 &  J131258.27+311531.0 & 198.24282 &  31.25862 & 0.0204 & $ 9.56\pm0.14$ & 10.81 &  7.36 &  2.02 &  1 & $-0.10\pm0.06$ &  8.55 &  1 &  1 \\
 JINGLE22 &  J130329.08+263301.7 & 195.87117 &  26.55050 & 0.0221 & $10.56\pm0.01$ &  7.65 &  8.93 &  3.09 &  1 & $-0.11\pm0.02$ &  8.67 &  2 &  2 \\
 JINGLE23 &  J120018.00+001741.9 & 180.07501 &   0.29499 & 0.0207 & $10.33\pm0.07$ &  7.59 &  8.33 &  2.56 &  1 & $-0.16\pm0.02$ &  8.92 &  1 &  2 \\
 JINGLE24 &  J121520.15-002352.9 & 183.83397 &  -0.39805 & 0.0208 & $ 9.10\pm0.12$ &  4.58 &  7.98 &  2.69 &  1 & $-0.55\pm0.06$ &  8.57 &  1 &  2 \\
 JINGLE25 &  J130636.39+275222.6 & 196.65164 &  27.87295 & 0.0209 & $10.12\pm0.02$ &  6.14 &  8.40 &  2.93 &  1 & $ 0.05\pm0.02$ &  8.72 &  1 &  2 \\
 JINGLE26 &  J130916.08+292203.5 & 197.31702 &  29.36765 & 0.0209 & $ 9.61\pm0.01$ &  4.24 &  8.26 &  2.56 &  1 & $ 0.07\pm0.00$ &  8.67 &  1 &  1 \\
 JINGLE27 &  J121552.50+002402.5 & 183.96875 &   0.40070 & 0.0210 & $10.49\pm0.05$ &  9.93 &  8.73 &  2.64 &  1 & $-0.06\pm0.04$ &  8.87 &  1 &  1 \\
 JINGLE28 &  J131047.64+294235.6 & 197.69852 &  29.70990 & 0.0212 & $ 9.91\pm0.13$ &  8.61 &  8.37 &  2.18 &  1 & $ 0.02\pm0.03$ &  8.85 &  1 &  1 \\
 JINGLE29 &  J115846.24-012757.0 & 179.69268 &  -1.46584 & 0.0214 & $ 9.78\pm0.00$ &  3.09 &  8.33 &  2.58 &  1 & $-0.33\pm0.00$ &  8.84 &  1 &  3 \\
 JINGLE30 &  J130558.70+252756.4 & 196.49460 &  25.46569 & 0.0218 & $ 9.76\pm0.14$ &  3.56 &  8.75 &  2.74 &  1 & $-0.26\pm0.09$ &  8.85 &  1 &  2 \\
 JINGLE31 &  J131502.15+280210.9 & 198.75896 &  28.03636 & 0.0218 & $ 9.55\pm0.07$ &  3.20 &  8.47 &  2.50 &  1 & $-0.20\pm0.08$ &  8.76 &  1 &  1 \\
 JINGLE32 &  J125809.99+242056.1 & 194.54164 &  24.34893 & 0.0226 & $ 9.49\pm0.00$ &  2.21 &  8.75 &  3.09 &  1 & $-0.17\pm0.03$ &  8.45 &  1 &  2 \\
 JINGLE33 &  J130945.77+283716.3 & 197.44071 &  28.62121 & 0.0226 & $ 9.36\pm0.00$ &  3.33 &  8.55 &  2.37 &  1 & $-0.74\pm0.00$ &  8.78 &  1 &  3 \\
 JINGLE34 &  J132703.18+305836.6 & 201.76327 &  30.97685 & 0.0227 & $ 9.57\pm0.06$ &  6.21 &  8.22 &  2.55 &  1 & $ 0.10\pm0.02$ &  8.74 &  1 &  1 \\
 JINGLE35 &  J131958.31+281449.3 & 199.99299 &  28.24704 & 0.0227 & $10.19\pm0.08$ &  4.21 &  8.82 &  2.56 &  1 & $-0.05\pm0.02$ &  8.84 &  1 &  2 \\
 JINGLE36 &  J130851.54+283745.4 & 197.21477 &  28.62928 & 0.0227 & $ 9.51\pm0.06$ &  4.52 &  8.13 &  2.56 &  1 & $-0.34\pm0.02$ &  8.63 &  1 &  3 \\
 JINGLE37 &  J131508.21+302413.5 & 198.78423 &  30.40377 & 0.0232 & $10.50\pm0.00$ &  4.99 &  8.96 &  2.93 &  1 & $ 0.06\pm0.00$ &  8.83 &  1 &  2 \\
 JINGLE38 &  J131928.01+274456.2 & 199.86671 &  27.74897 & 0.0232 & $ 9.76\pm0.07$ &  2.06 &  9.03 &  2.82 &  1 & $-0.21\pm0.18$ &  8.70 &  1 &  3 \\
 JINGLE39 &  J132638.85+270223.4 & 201.66187 &  27.03984 & 0.0233 & $ 9.56\pm0.11$ &  7.32 &  8.05 &  2.31 &  1 & $-0.24\pm0.05$ &  8.60 &  1 &  3 \\
 JINGLE40 &  J131745.18+273411.5 & 199.43827 &  27.56987 & 0.0233 & $10.59\pm0.03$ &  7.86 &  9.04 &  3.24 &  1 & $ 0.15\pm0.02$ &  8.74 &  2 &  2 \\
 JINGLE41 &  J130617.29+290347.4 & 196.57206 &  29.06318 & 0.0234 & $10.83\pm0.03$ & 10.14 &  9.19 &  2.66 &  1 & $ 0.45\pm0.02$ &  8.76 &  3 &  2 \\
 JINGLE42 &  J132643.48+303024.0 & 201.68117 &  30.50668 & 0.0236 & $ 9.86\pm0.11$ &  3.92 &  8.63 &  2.51 &  1 & $-0.34\pm0.03$ &  8.63 &  2 &  1 \\
 JINGLE43 &  J133457.27+340238.7 & 203.73864 &  34.04408 & 0.0236 & $10.59\pm0.01$ & 12.06 &  8.58 &  2.24 &  1 & $ 1.49\pm0.00$ &  8.88 &  1 &  3 \\
 JINGLE44 &  J130143.37+290240.7 & 195.43072 &  29.04466 & 0.0237 & $10.63\pm0.11$ & 11.14 &  8.89 &  2.91 &  1 & $ 0.11\pm0.05$ &  8.73 &  1 &  3 \\
 JINGLE45 &  J130831.57+244202.7 & 197.13158 &  24.70076 & 0.0238 & $10.80\pm0.07$ & 13.01 &  8.87 &  3.09 &  1 & $ 0.09\pm0.05$ &  8.84 &  2 &  1 \\ 
\hline
\end{tabular}
\end{table}
\end{landscape}

\appendix
\section{Spectral energy distributions}
\label{SEDsection}

We present for each of the \ntot\ galaxies in the JINGLE sample the spectral energy distributions obtained from the CAAPR photometric catalog (see Section \ref{caapr} for details). Each SED is accompanied by the best fitting models obtained with {\sc MAGPHYS}, and with the templates of \citet{ce01} and \citet{mullaney11}. Details of the modeling and of these specific templates are given in Section \ref{SFRsection}. 

\begin{figure*}
    \centering
    \includegraphics[width=7in]{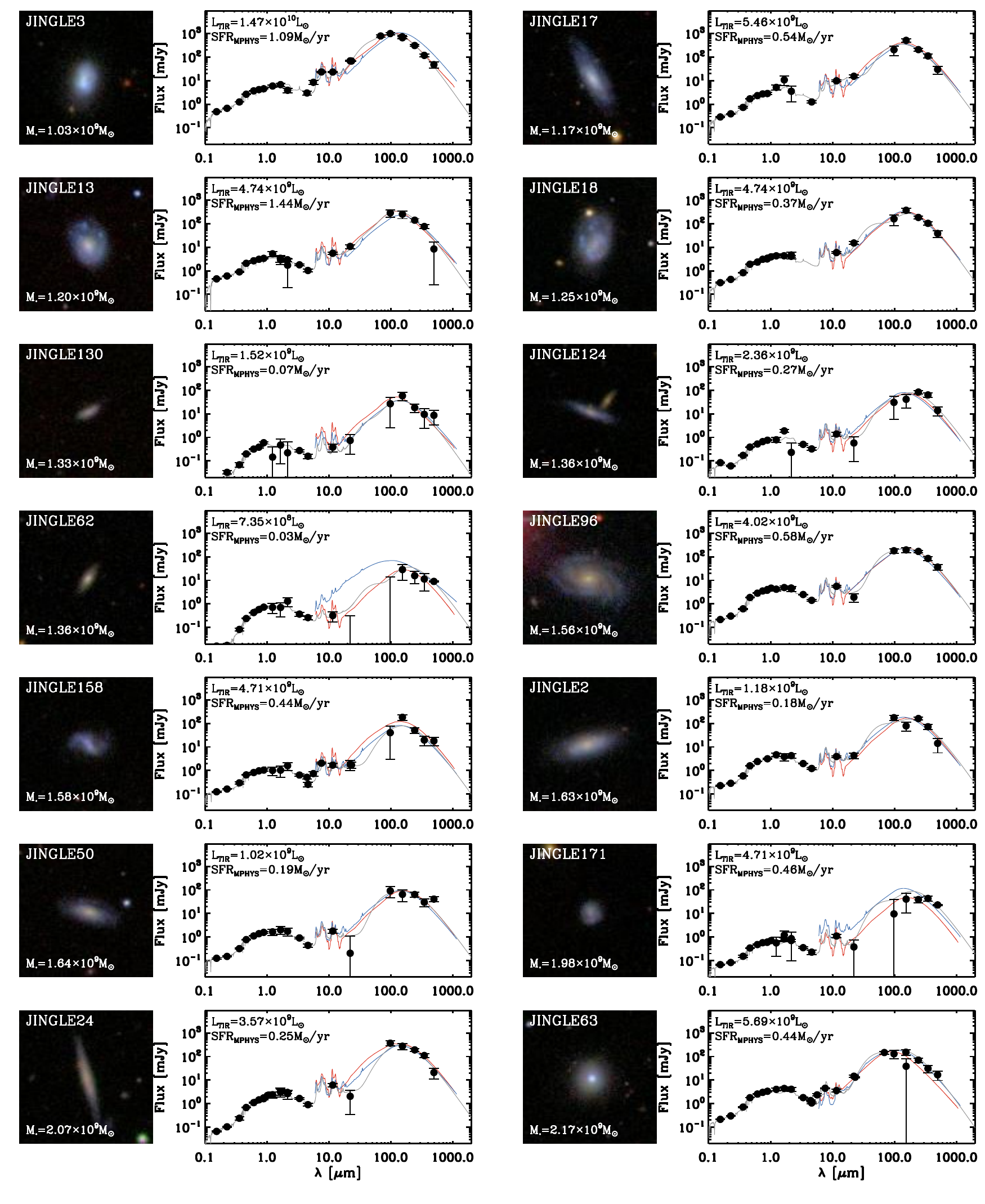}
    \caption{For each JINGLE galaxy, {\it left:} SDSS image, 1\arcmin$\times$1\arcmin, {\it right:} Full SED from CAAPR as well as the {\sc MAGPHYS} model (gray line) and modeling of the FIR SED using the templates of CE01 (red line) and JRM (blue line). These SEDs, and the fits to them, do not include the JCMT measurements.}
    \label{SED1}
\end{figure*}

\begin{figure*}
    \centering
    \includegraphics[width=7in]{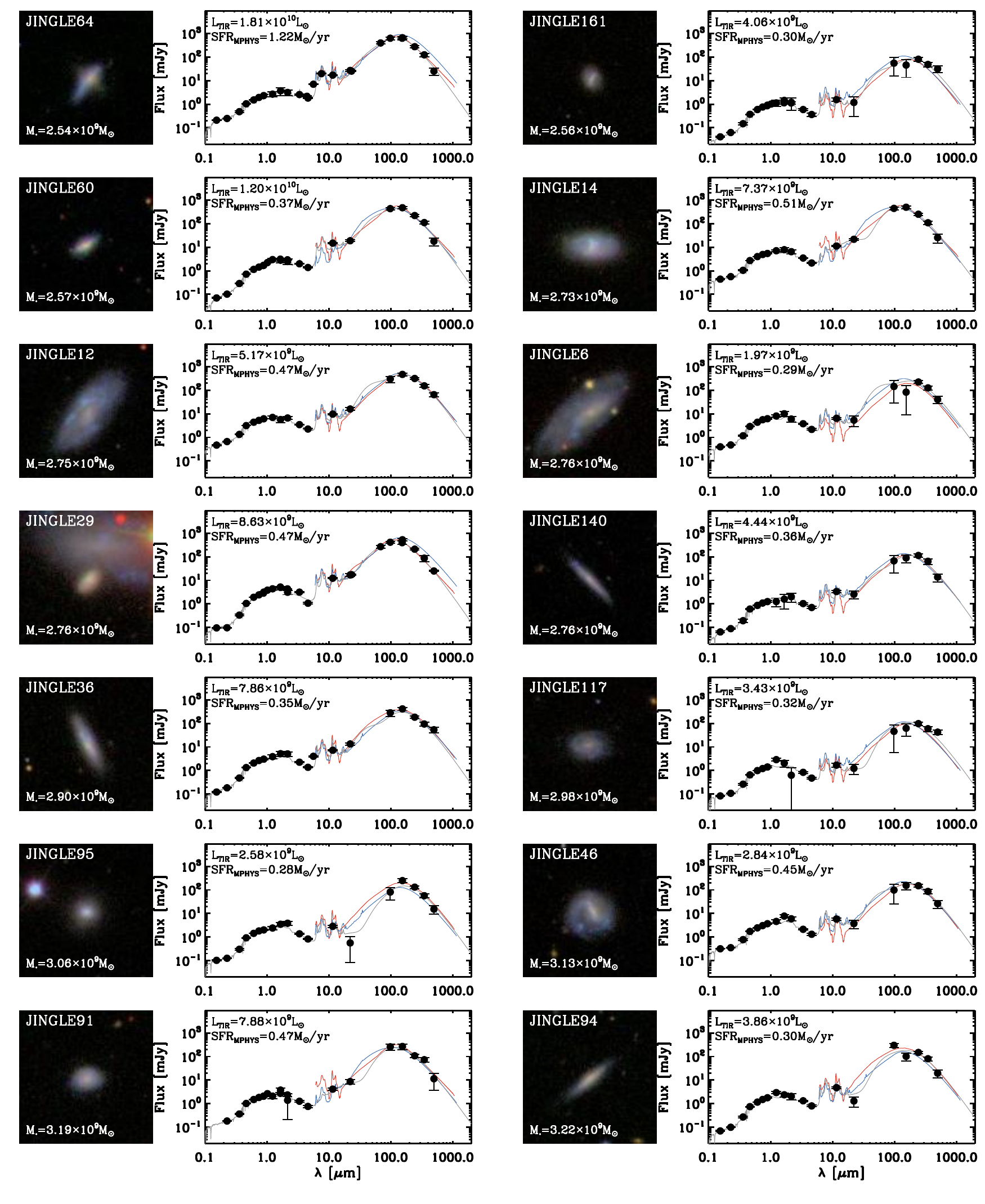}
    \caption{Continued from Fig. \ref{SED1}.}
\end{figure*}

\begin{figure*}
    \centering
    \includegraphics[width=7in]{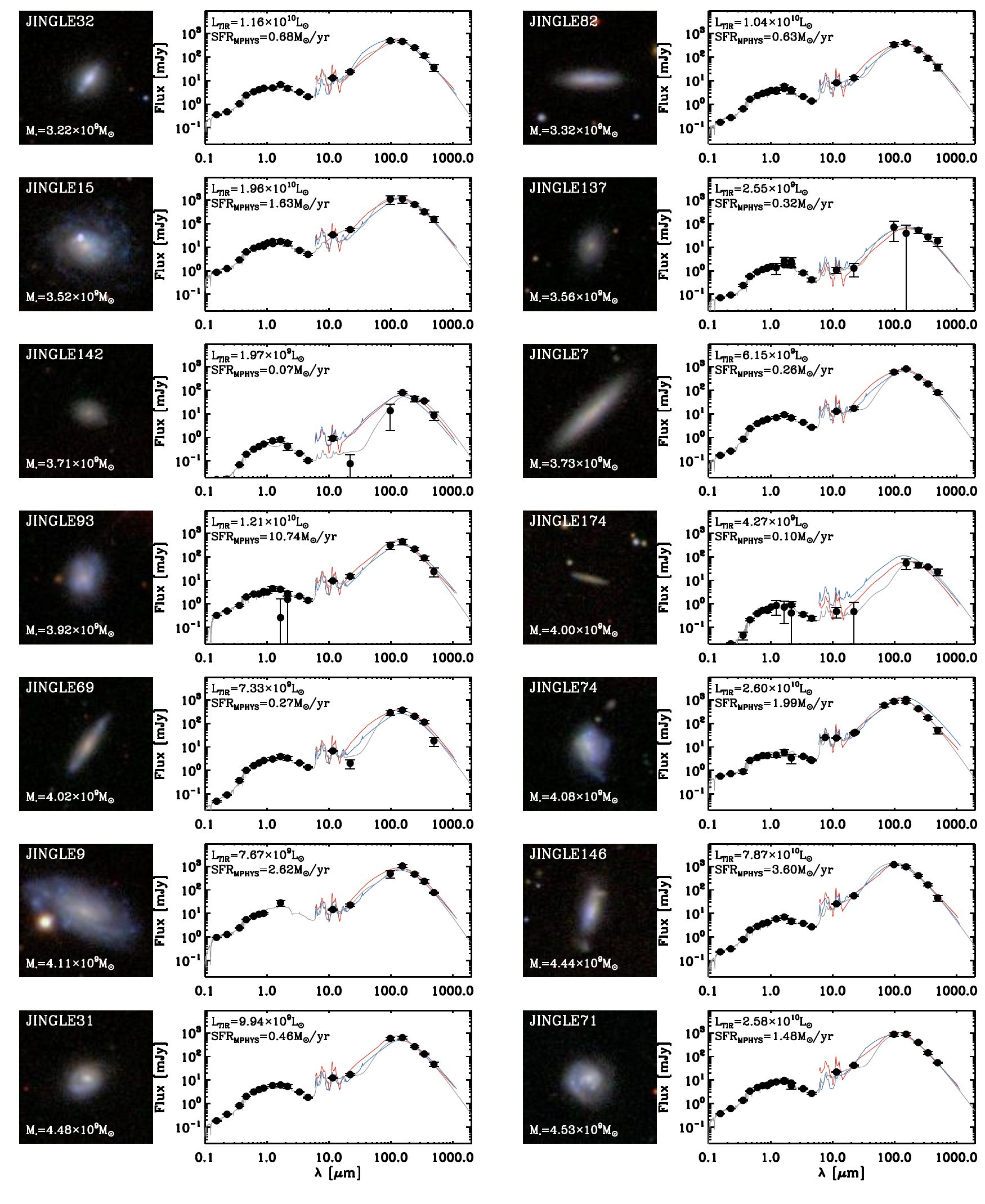}
    \caption{Continued from Fig. \ref{SED1}. The remaining 10 pages of this figure are available online. }
\end{figure*}

\label{lastpage}
\end{document}